# Encounter-based worms: Analysis and Defense

Sapon Tanachaiwiwat
Department of Electrical Engineering
University of Southern California, CA
tanachai@usc.edu

Ahmed Helmy
Computer and Information Science and Engineering
University of Florida, FL
helmy@ufl.edu

Abstract— Encounter-based network is a frequently-disconnected wireless ad-hoc network requiring immediate neighbors to store and forward aggregated data for information disseminations. Using traditional approaches such as gateways or firewalls for deterring worm propagation in encounter-based networks is inappropriate. We propose the worm interaction approach that relies upon automated beneficial worm generation aiming to alleviate problems of worm propagations in such networks. To understand the dynamic of worm interactions and its performance, we mathematically model worm interactions based on major worm interaction factors including worm interaction types, network characteristics, and node characteristics using ordinary differential equations and analyze their effects on our proposed metrics. We validate our proposed model using extensive synthetic and trace-driven simulations. We find that, all worm interaction factors significantly affect the pattern of worm propagations. For example, immunization linearly decreases the infection of susceptible nodes while *on-off* behavior only impacts the duration of infection. Using realistic mobile network measurements, we find that encounters are "bursty", multi-group and non-uniform. The trends from the trace-driven simulations are consistent with the model, in general. Immunization and timely deployment seem to be the most effective to counter the worm attacks in such scenarios while cooperation may help in a specific case. These findings provide insight that we hope would aid to develop counter-worm protocols in future encounter-based networks.

## I. INTRODUCTION

An encounter-based network is a frequently-disconnected wireless ad-hoc networks requiring close proximity of neighbors, i.e., encounter, to disseminate information. Hence, we call this the "encounter-based network" which can be considered as a terrestrial delay-and-disruptive-tolerant network. It is an emerging technology that is suitable for applications in highly dynamic wireless networks.

Most previous work on worm propagation has focused on modeling single worm type in well-connected wired network. However, many new worms are targeting wireless mobile phones. The characteristics of worms in mobile networks are different from random-scan network worms. Worm propagations in mobile networks depend heavily on user encounter patterns. Many of those worms rely on Bluetooth to broadcast their replications to vulnerable phones, e.g., Cabir and ComWar.M [10, 13]. Since Bluetooth radios have very short range around 10-100 meters, the worms need neighbors in close proximity to spread out their replications. Hence, we call this "encounter-based worms". This worm spreading pattern is very similar to spread of packet replications in delay tolerant networks [15, 17], i.e., flooding the copies of messages to all close neighbors. An earlier study in encounter-based networks actually used the term "*epidemic routing*" [15] to describe the similarity of this routing protocol to disease spreading. Using traditional approaches such as gateways or firewalls for deterring worm propagation in encounter-based networks is inappropriate. Because this type of network is highly dynamic and has no specific boundary, a fully distributed counter-worm mechanism is needed. We propose to investigate the worm interaction approach that relies upon automated beneficial worm generation [1]. This approach uses an automatic generated beneficial worm to terminate malicious worms and patch vulnerable nodes.

Our work is motivated by wars of Internet worms such as the war between NetSky, Bagle and MyDoom [13]. This scenario is described as "worm interactions" in which one or multiple type of worm terminates or patches other types of worms.

In this paper, we mathematically model worm interactions based on three major worm interaction factors including worm interaction types [11], network characteristics and node characteristics [12]. Worm interaction types in our model are *aggressive one-sided*, *conservative one-sided*, *aggressive two-sided*. The variation of these worm interaction types can also be created from our model.

There are many important node characteristics to be considered, but we focus only a fundamental subset including cooperation, immunization, *on-off* behavior and delay. We shall show that these are key node characteristics for worm propagation in encounter-based networks. Other characteristics such as trust between users, battery life, energy consumption, and buffer capacity are subject to further study and are out of scope of this paper.

The majority of routing studies in encounter-based networks usually assume ideal node characteristics including full node cooperation and *always-on* behavior. However, in realistic scenarios, nodes do not always cooperate with others and may be *off* most of the time [5]. In worm propagation studies, many works also assume all nodes to be susceptible (i.e., not immune) to worm infection. An *immune* node does not cooperate with infected nodes and is not infected. To investigate more realistic scenarios, we propose to study the mobile node characteristics and analyze the impact of cooperation, immunization and *on-off* behavior on the worm interactions. Cooperation and *on-off* behavior are expected to have impact on the timing of infection. Intuitively, cooperation makes the network more susceptible to worm attacks. Immunization, however, may help reduce overall infection level. This paper examines the validity of these expectations, using the overall infection level and timing of infection as metrics (see Section III.*C*).

We consider several important network characteristics such as node sizes, contact rate, group behaviors and batch arrival. Using realistic mobile network measurements, we find that encounters are *"bursty", multi-group and non-uniform*.

Most worm propagation studies only focus on instantaneous number of infected nodes as a metric. We feel that additional systematic metrics are needed to study worm response mechanisms. We utilize new metrics including total prey-infected nodes, maximum prey-infected nodes, total prey lifespan, average individual prey lifespan, time to secure all nodes, and time to remove all preys to quantify the effectiveness of worm interaction.

In this paper, we try to answer following questions: How can we model this *war of the worms* systemically based on worm interaction factors including worm interaction types, node characteristics and network characteristics? What type of worm interaction, conditions of network and node characteristics can alleviate the level of worm infection? How do worms interact in realistic mobility scenario? This worm interaction model can be extended to support more complicated current and future worm interactions in encounter-based networks.

Our main contributions in this paper are our proposed new *Worm Interaction Model* focusing on worm interaction types, network characteristics and node characteristics in encounter-based networks. We also use new metrics to quantify the effectiveness of worm interactions, and our proposed metrics are applicable to study any worm response mechanism. We also provide the first study of worm propagation based on real mobile measurements.

Following is an outline of the rest of the paper. We discuss related work in Section II. Then, in Section III, we explain the basic definitions of our model, the metrics, worm interaction types, network characteristics, node characteristics and the general model. Then we analyze and evaluate worm interactions in both uniform and realistic encounter networks in Section IV. In Section V, we conclude our work and discuss the future work.

## II RELATED WORK

Worm-like message propagation or epidemic routing has been studied for delay tolerant network applications [11, 13, 15]. As in worm propagation, a sender in this routing protocol spreads messages to all nodes in close proximity, and those nodes repeatedly spread the copies of messages until the messages reach a destination, similarly to generic flooding but without producing redundant messages. Performance modeling for epidemic routing in delay tolerant networks [13] based on ordinary differential equations (ODE) is proposed to evaluate the delivery delay, loss probability and power consumption. Also the concept of anti-packet is proposed to stop unnecessary overhead from forwarding extra packets copies after the destination has received the packets. This can be considered as a special case of non-zero delay of aggressive one-sided interaction (see Section III.*B*) which we consider in our model.

Epidemic models, a set of ODEs, were used to describe the contagious disease spread including *SI, SIS, SIR SIRS, SEIR* and *SEIRS* models [4, 10] in which *S, I, E, R* stand for Susceptible, Infected, Exposed and Recovered states, respectively. There's an analogy between computer worm infection and disease spread in that both depend on node's state and encounter pattern. For Internet worms, several worm propagation models have been investigated in earlier work [2, 6, 8, 18]. Few works [1, 9, 11] consider worm interaction among different worm types. Our work, by contrast, focuses on understanding of how we can systemically categorize and model worm propagation based on worm interaction types, network characteristics and node characteristics in encounter-based networks.

In [1], the authors suggest modifying existing worms such as Code Red, Slammer and Blaster to terminate the original worm types. In this paper, we model this as aggressive one-sided worm interaction. Other active defenses, such as automatic patching, are also investigated in [16]. Their work assumes a patch server and overlay network architecture for Internet defense. We provide a mathematical model that can explain the behavior of automatically-generated beneficial worm and automatic patch distribution using one-sided worm interaction in encounter-based networks. Effect of Immunization on Internet worms was modeled in [8] based on the *SIR* model.

## III. WORM INTERACTION MODEL

We aim to build a fundamental worm propagation model that captures worm interaction as a key factor in uniform encounter-based networks. Furthermore, our proposed model addresses and analyzes dynamics of susceptible and infected nodes over the course of time.

Because the constant removal rate in basic *SIR* model and its variance [7, 14] cannot directly portray such interactions impact on multi-type worm propagations, our model builds upon and extends beyond the conventional epidemic model to accommodate the notion of interaction.

Basic operation of a worm is to find susceptible nodes to be infected and the main goal of attackers is to have their worms infect the largest amount of nodes in the least amount of time, and if possible, undetected by antivirus or intrusion detection systems. Our beneficial worm, on the other hand, aims to eliminate opposing worms or limit the scope of opposing worms' infection. We want to investigate the worm propagation caused by various types of interactions as well as network characteristics and node characteristics.

### *A. Definitions*

#### a. Predator-Prey Relationships

For every worm interaction type, there are two basic characters: Predator and Prey. The ***Predator***, in our case the beneficial worm, is a worm that terminates and patches against another worm. The ***Prey***, in our case the malicious worm, is a worm that is terminated or patched by another worm.

A predator can also be a prey at the same time for some other type of worm. Predator can *vaccinate* a susceptible node, i.e., infect the susceptible node (vaccinated nodes become predator-infected nodes) and apply a patch afterwards to prevent the nodes from prey infection. Manual vaccination, however, is performed by a user or an administrator by applying patches to susceptible nodes.

A *termination* refers to the removal of prey from infected nodes by predator; and such action causes prey-infected nodes to become predator-infected nodes. The removal by a user or an administrator, however, is referred to as *manual removal.*

We choose to use two generic types of interacting worms, *A* and *B*, as our basis throughout the paper. *A* and *B* can assume the role of predator or prey depending on the type of interactions.

#### b. Contact Rate

Contact rate is the frequency of encounter for pairs of nodes, where an encounter occurs when the 2 nodes are within radio range. We assume a uniform contact rate for all pairs of nodes and their encounter behavior does not directly impact each other and both predator and prey share the same set of susceptible nodes. We assume that in one encounter, worm is successfully transferred from one node to another.

**c. Metrics**

To gain insight and better quantify the effectiveness of worm interaction, we propose to use the following metrics:

**(1) Total Prey-infected Nodes (*TI*):** the number of nodes ever infected by prey.

**(2) Maximum Prey-infected Nodes (*MI*):** the peak of instantaneous number of prey-infected nodes where $I_A(0) \le MI \le TI$ .

**(3) Total Prey Lifespan (*TL*):** the sum of time of individual nodes ever infected by prey. It can be interpreted as the total damage by prey.

**(4) Average Individual Prey Lifespan (*AL*):** the average lifespan of individual prey-infected nodes where $AL \le TL$ .

**(5) Time to Secure All Nodes (*TA*):** the time required for predator to infect all susceptible and prey nodes. Its inverse can be interpreted as average predator infection rate.

**(6) Time to Remove All Preys (*TR*):** the time required for predator to terminate all preys where $TR \le TA$ . Its inverse can be interpreted as prey termination rate.

*TI* and *MI* are indicators of the level of prey infection, *TL* and *AL* are the indicators of the duration of prey infection and *TA* and *TR* are the indicators of protection and recovery rate, respectively. Our goal is to find the conditions to *minimize* these metrics based on worm interaction factors of which details are discussed next.

***B. Worm interaction factors***

Our model considers three major factors that can significantly impact the worm interactions: worm interaction types, network characteristics, and node characteristics. Worm can behave differently based on types of interactions (or their behaviors): aggressive one-sided interaction, conservative one-sided interaction or aggressive two-sided interaction [11]. In addition, underlying network characteristics including node size, contact rate, group behaviors and batch arrivals are the key of worm propagation. Finally, node characteristics: cooperation, immunization, *on-off* behaviors and delay, can significantly affect the worm interaction patterns. We start by explaining each individual worm interaction factors before we show our model that addressing all of these factors.

**a. Worm interaction types**

When there is a prey, *A*, and a predator, *B*, we consider this as a one-sided interaction. If both *A* and *B* are predators, it is denoted as a two-sided interaction. For ideal scenario, the predator wants to terminate its prey as much as possible as well as prevent its preys from infection and re-infection. To satisfy that requirement, the predator requires a patch or a false signature of its prey.

There are three types of interactions considered: aggressive one-sided, conservative one-sided and aggressive two-sided. Followings are their descriptions.

**(1) Aggressive one-sided interaction:** In this interaction type, a beneficial worm, predator has the capability to terminate and patch a malicious worm, prey, as well as vaccinate susceptible nodes. Simplified interaction between the Internet worms, e.g., Welchia and Blaster can be represented by this model.

**(2) Conservative one-sided interaction:** In a conservative interaction, a predator has the capability to terminate a prey but does *not* vaccinate susceptible nodes. Hence the predator-infected nodes change depends solely on population of the prey-infected nodes.

**(3) Aggressive two-sided interaction:** In this interaction type, both worms assume the roles of predator and prey simultaneously. We would simply call $A$ as *predator A* and $B$ as *predator B*. Predator $B$ is capable of vaccinating susceptible nodes but unable to remove a predator $A$ from predator $A$'s infected nodes because it is blocked by predator $A$. Both predator A and B blocks each other. In automated patching systems [16], their worm-like patch distribution falls into this category. The automated patching assumes that each worm patches its own node to prevent infection from other worm is closely related to this model.

According to above worm interaction types, *TI, MI, TL, AL, TA* and *TR in* aggressive one-sided interaction are expected to be the lowest among those of all interaction types. In conservative one-sided interaction, because only once-infected-by-prey nodes can be infected by predator, hence $TA = \infty$. Similarly, for aggressive two-sided interaction, predator cannot terminate prey, hence $TL = AL = TA = TR = \infty$.

**b. Network characteristics**

Network characteristics represent the characteristics of the encounter-based networks. We particularly focus on node sizes, contact rate, group behaviors and batch arrival. The other related characteristics including clustering coefficient, average hop counts are subject to further study.

**(1) Contact rate:** Contact rate ($\beta$) is one of the most important factors to decide the characteristics of worm interaction. We investigate the relationships between $\beta$ and our proposed metrics here in this section. Because contact rate is the frequency of a pair of nodes encounter each other, increasing the contact rate causes every node to encounter each other more frequently, i.e., the time between consecutive encounters will be reduced. Hence, we expect that the metrics relating to times including *TL*, *AL*, *TA* and *TR* to be reduced. However, because as prey and predator share the same contact rate, *TI* and *MI* should not be different even when contact rates are changed. In other words, if prey infects other susceptible nodes faster, predator also terminates and patches faster as well.

**(2) Node size:** With the same number of initial predator and initial prey-infected nodes, when node size (*N)* is changed with fixed $\beta$, this implies the decrease of time between consecutive encounter of any node to any node. Similarly as we expect from contact rate, varying node sizes can have significant impact on *TL*, *AL*, *TA* and *TR*.

**(3) Group behavior**: Multi-group encounters, of which group is classified by their encounter patterns and contact rates, are expected in encounter-based networks. For two-group modeling, we need 3 different contact rates: two intra-contact rates for encounters within each group, and one inter-contact rate for encounters between groups. For *n* groups, we need *n* intra-contact rates and $\binom{n}{2}$ inter-contact rates. Effects of group sizes, contact rates of the individual group and between groups are investigated.

**(4) Batch arrival:** Nodes may join the networks simultaneously as a "batch arrival". It can be modeled as the "birth" of the population. We assume that those nodes enter the network only as susceptible nodes. Note that for infected nodes that temporarily leave and then join the network, we would not consider this case as a batch arrival. We discuss and investigate the effect of realistic batch arrival in Section IV.

**c. Node characteristics**

Each node may have different characteristics because of differences in user's usage strategies, daily-life activities or level of security technology and awareness. Four important node characteristics corresponding to this worm interaction factor are addressed including cooperation, immunization, *on-off* behavior and delay. We assume these node characteristics are persistent through out the life time of the networks.

**(1) Cooperation:** Cooperation is the willingness of node to forward messages (worms) for other nodes. The opposite characteristic is known as selfishness. Intuitively, cooperation may seem to make the network more vulnerable. However, unlike immunization, cooperation is expected to equally slow down both prey and predator propagations. Hence, the effect of cooperation is hard to anticipate.

**(2) Immunization:** Not all nodes are susceptible to the prey either because of their heterogeneous operating systems and their differences of promptness to remove the vulnerability from their machines. Hence partial of nodes can be immune to prey and will slow down the overall prey infection. It is expected to improve the overall targeted metrics that we mention earlier because immune nodes still help forward predator to other nodes. It is expected to have no positive impact on *TA* but reduce *TR* simply because of less number of nodes to be removed.

**(3) *On-off* behavior:** A node is able to accept or forward the packet based on the *on-off* characteristics. In reality, devices are "*on*" or active only a fraction of the time. Activity may be related to mobility. For instance, a mobile phone is usually *on*, while lap top is unlikely to be mobile while *on*[1]. We model the transition from *on* to *off*, and vice versa, probabilistically. The probability is determined at the beginning of each time interval. Hence the contact rate is expected to be proportionally reduced according to the probability that the node cannot forward or accept the packets because of *on-off* status.

**(4) Delay:** Initial prey-infected nodes and initial predator-infected nodes may start their infections in the networks at different times (depending on prey timers or security architecture of predator). The gap between those times can be significant. If initial prey-infected nodes start infecting susceptible nodes in the network earlier than initial predator-infected nodes starts vaccination and termination, we can expect the increase of *TI, MI, AL*, *TA*, *TL* and *TR*, and the opposites results are expected if the order of their start times are reversed.

### *C.* General Worm Interaction Model

Assume that there are *g* groups in the network. Let $\beta_{nm}$ is the contact rate between member of group *n* and group *m* ($\beta_{nn}$ is the contact rate within group *n*), $S_n$ is the number of susceptible nodes of group *n* (at time *t*) where $1 \le m, n \le g$. Let *c* be the fraction of $N_n$ that are willing to be *cooperative* where $0 \le c \le 1$ and $N_n$ is the total number of nodes in the networks for group *n*. Let *i* be the fraction of cooperative nodes that are *immune* to prey where $0 \le i \le 1$. Let $I_{An}$ and $I_{Bn}$ be the number of prey-infected nodes and predator-infected nodes for group *n*, respectively. We assume that initial predator-infected and initial prey-infected nodes (*t*=0)are cooperative then the number of susceptible nodes for both prey and predator is $S^*_n$ where $S^*_n(0) = c(1-i)N_n - I_{An}(0)$ for group *n* and number of susceptible nodes for predator only is $S'_n$, where $S'_n(0) = ciN_n - I_{Bn}(0)$ for group *n*. Note that $N_n = S^*_n + S'_n + I_{An} + I_{Bn}$ and $S_n = S^*_n + S'_n$. We define the probability of "*on*" behavior as *p* and "*off*" behavior as 1-*p* where $0 \le p \le 1$. Hence contact rate between group *n* and *m* for both predator and prey is $p\beta_{nm}$. Let *d* be the delay between the initial prey-infected node(s) and the initial predator-infected node(s) (assume all initial predator-infected (prey-infected) nodes start infection at the same time) then $I_{An}(t) \geq 1$ iif $t \geq 0$ and $I_{Bn}(t) \geq 1$ iif $t \geq d$. For simplicity and brevity, let us assume that number of groups in the network is 2. Fig.1a shows the state diagram of our model.

[1] This is observed from measurements [15] and is captured in our study using trace-driven simulations.

Let $K_{S*_1 I_{A1} I_{A2}}$, $K_{S*_2 I_{A1} I_{A2}}$, $K_{S*_1 I_{B1} I_{B2}}$, $K_{S*_2 I_{B1} I_{B2}}$, $K_{S'_1 I_{B1} I_{B2}}$, $K_{S'_2 I_{B1} I_{B2}}$, $K_{I_{A1} I_{B1} I_{B2}}$ and $K_{I_{A2} I_{B1} I_{B2}}$ be the state transition indicator from $S*_1$ to either $I_{A1}$ or $I_{A2}$, where $K_{S*_1 I_{A1} I_{A2}} \in \{0,1\}$, from $S*_2$ to either $I_{A1}$ or $I_{A2}$ where $K_{S*_2 I_{A1} I_{A2}} \in \{0,1\}$, from $S*_1$ to either $I_{B1}$ or $I_{B2}$ where $K_{S*_1 I_{B1} I_{B2}} \in \{0,1\}$, from $S*_2$ to either $I_{B1}$ or $I_{B2}$ where $K_{S*_2 I_{B1} I_{B2}} \in \{0,1\}$, from $S'_1$ to either $I_{B1}$ or $I_{B2}$ where $K_{S'_1 I_{B1} I_{B2}} \in \{0,1\}$, from $S'_2$ to either $I_{B1}$ or $I_{B2}$ where $K_{S'_2 I_{B1} I_{B2}} \in \{0,1\}$, from $I_{A1}$ to either $I_{B1}$ or $I_{B2}$ where $K_{I_{A1} I_{B1} I_{B2}} \in \{0,1\}$, and from $I_{A2}$ to either $I_{B1}$ or $I_{B2}$ where $K_{I_{A2} I_{B1} I_{B2}} \in \{0,1\}$ ,respectively. Let $\alpha$ be the rate that prey-infected or predator-infected nodes become susceptible again ($\alpha$ can also be different between prey and predator). The state transition indicators and $\alpha$ are used to identify types of worm interactions. Let $\gamma$ be the manual removal rate and $\gamma_S$ be the manual vaccination.

For the aggressive one-sided interaction, $K_{S*_1 I_{A1} I_{A2}} = K_{S*_2 I_{A1} I_{A2}} = K_{S*_1 I_{B1} I_{B2}} = K_{S*_1 I_{B1} I_{B2}} = K_{S'_1 I_{B1} I_{B2}} = K_{S'_2 I_{B1} I_{B2}} = K_{I_{A1} I_{B1} I_{B2}} = K_{I_{A2} I_{B1} I_{B2}} = 1$ and $\alpha = 0$, for the conservative one-sided interaction, $K_{S*_1 I_{A1} I_{A2}} = K_{S*_2 I_{A1} I_{A2}} = K_{I_{A1} I_{B1} I_{B2}} = K_{I_{A2} I_{B1} I_{B2}} = 1$, $K_{S*_1 I_{B1} I_{B2}} = K_{S*_1 I_{B1} I_{B2}} = K_{S'_1 I_{B1} I_{B2}} = K_{S'_2 I_{B1} I_{B2}} = 0$ and $\alpha = 0$, for the aggressive two-sided interaction, $K_{S*_1 I_{A1} I_{A2}} = K_{S*_2 I_{A1} I_{A2}} = K_{S*_1 I_{B1} I_{B2}} = K_{S*_1 I_{B1} I_{B2}} = K_{S'_1 I_{B1} I_{B2}} = K_{S'_2 I_{B1} I_{B2}} = 1$, $K_{I_{A1} I_{B1} I_{B2}} = K_{I_{A2} I_{B1} I_{B2}} = 0$ and $\alpha = 0$.

Let $\lambda_{S*_1 S*_2}, \lambda_{S*_2 S*_1}, \lambda_{S'_1 S'_2}, \lambda_{S'_2 S'_1}, \lambda_{I_{A1} I_{A2}}, \lambda_{I_{A2} I_{A1}}, \lambda_{I_{B1} I_{B2}}$ and $\lambda_{I_{B2} I_{B1}}$ be the group transition rates from $S*_1$ to $S*_2$, $S*_2$ to $S*_1$, $S'_1$ to $S'_2$, $S'_2$ to $S'_1$, $I_{A1}$ to $I_{A2}$, $I_{A2}$ to $I_{A1}$, $I_{B1}$ to $I_{B2}$, and $I_{B2}$ to $I_{B1}$ ,respectively. Let $\Delta_{S*_1}$, $\Delta_{S*_2}$, $\Delta_{S'_1}$, and $\Delta_{S'_2}$ be the batch arrival rates for $S*_1$, $S*_2$, $S'_1$ and $S'_2$, respectively.

Susceptible nodes' decrease rate is determined by manual vaccination and the contact of susceptible nodes with the prey-infected nodes (from the same or different group) causing the prey infection or with the predator-infected nodes (from the same or different group) causing the vaccination. On the other hand, the re-susceptible (infected nodes become susceptible again[2]) rate causes the increase for susceptible nodes. In addition, the number of susceptible nodes within each group can be changed due to the group transitions and batch arrival. Hence, the susceptible rates of group 1 and 2 are

$$\frac{dS*_1}{dt} = -pS*_1 \left(K_{S*_1 I_{A1} I_{A2}}(\beta_{11} I_{A1} + \beta_{12} I_{A2}) + K_{S*_1 I_{B1} I_{B2}}(\beta_{11} I_{B1} + \beta_{12} I_{B2})\right) + (\lambda_{S*_2 S*_1} S*_2 - \lambda_{S*_1 S*_2} S*_1) - \gamma_S S*_1 + \alpha(I_{A1} + (1-i) I_{B1}) + \Delta_{S*_1} \quad \text{(2-a)}$$

$$\frac{dS*_2}{dt} = -pS*_2 \left(K_{S*_2 I_{A1} I_{A2}}(\beta_{22} I_{A2} + \beta_{12} I_{A1}) + K_{S*_2 I_{B1} I_{B2}}(\beta_{22} I_{B2} + \beta_{12} I_{B1})\right) - (\lambda_{S*_2 S*_1} S*_2 - \lambda_{S*_1 S*_2} S*_1) - \gamma_S S*_2 + \alpha(I_{A2} + (1-i) I_{B2}) + \Delta_{S*_2} \quad \text{(2-b)}$$

$$\frac{dS'_1}{dt} = -pK_{S'_1 I_{B1} I_{B2}} S'_1 (\beta_{11} I_{B1} + \beta_{12} I_{B2}) + (\lambda_{S'_2 S'_1} S'_2 - \lambda_{S'_1 S'_2} S'_1) - \gamma_S S'_1 + \alpha i I_{B1} + \Delta_{S'_1} \quad \text{(2-c)}$$

[2] Some worms only reside in memory, and disappear after restart of computer

$$\frac{dS'_2}{dt} = -pK_{S'_2 I_{B1} I_{B2}} S'_2 (\beta_{22} I_{B2} + \beta_{12} I_{B1}) - (\lambda_{S'_2 S'_1} S'_2 - \lambda_{S'_1 S'_2} S'_1) - \gamma_S S'_2 + \alpha i I_{B2} + \Delta_{S'2} \quad \text{(2-d)}$$

Since the prey relies on susceptible nodes to expand its population, the increase of prey infection rate is determined by the contacts of susceptible nodes and prey-infected nodes. The decrease of prey infection rate is determined by prey termination caused by the contacts of prey-infected nodes and predator-infected nodes, manual removal rate and also the re-susceptible rate. The other factors such as group transition and batch arrival are also applied to the prey infection rate. Hence the prey infection rates for group 1 and 2 are

$$\frac{dI_{A1}}{dt} = p(K_{S^*_1 I_{A1} I_{A2}} S^*_1 (\beta_{11} I_{A1} + \beta_{12} I_{A2}) - K_{I_{A1} I_{B1} I_{B2}} I_{A1} (\beta_{11} I_{B1} + \beta_{12} I_{B2})) + (\lambda_{I_{A2} I_{A1}} I_{A2} - \lambda_{I_{A1} I_{A2}} I_{A1}) - (\alpha + \gamma) I_{A1} \quad \text{(3-a)}$$

$$\frac{dI_{A2}}{dt} = p(K_{S^*_2 I_{A1} I_{A2}} S^*_2 (\beta_{22} I_{A2} + \beta_{12} I_{A1}) - K_{I_{A2} I_{B1} I_{B2}} I_{A2} (\beta_{22} I_{B2} + \beta_{12} I_{B1})) - (\lambda_{I_{A2} I_{A1}} I_{A2} - \lambda_{I_{A1} I_{A2}} I_{A1}) - (\alpha + \gamma) I_{A2} \quad \text{(3-b)}$$

Because the predator can terminate its prey as well as vaccinate susceptible nodes, the increase of predator infection rate is determined by the contacts of predator with either the susceptible nodes or prey-infected nodes. The decreases of prey-infected nodes are caused by manual removal rate and re-susceptible rate. The predator infection rates for group 1 and 2 are

$$\frac{dI_{B1}}{dt} = p(\beta_{11} I_{B1} + \beta_{12} I_{B2})(K_{S^*_1 I_{B1} I_{B2}} S^*_1 + K_{S'_1 I_{B1} I_{B2}} S'_1 + K_{I_{A1} I_{B1} I_{B2}} I_{A1}) + (\lambda_{I_{B2} I_{B1}} I_{B2} - \lambda_{I_{B1} I_{B2}} I_{B1}) - (\alpha + \gamma) I_{B1} \quad \text{(4-a)}$$

$$\frac{dI_{B2}}{dt} = p(\beta_{22} I_{B2} + \beta_{12} I_{B1})(K_{S^*_2 I_{B1} I_{B2}} S^*_2 + K_{S'_1 I_{B1} I_{B2}} S'_2 + K_{I_{A2} I_{B1} I_{B2}} I_{A2}) - (\lambda_{I_{B2} I_{B1}} I_{B2} - \lambda_{I_{B1} I_{B2}} I_{B1}) - (\alpha + \gamma) I_{B2} \quad \text{(4-b)}$$

Finally, the increase of removed nodes is caused by manual vaccination of susceptible hosts and manual removal of prey-infected and predator-infected nodes.

$$\frac{dR}{dt} = \gamma_S (S^*_1 + S^*_2 + S'_1 + S'_2) + r(I_{A1} + I_{A2} + I_{B1} + I_{B2}) \quad \text{(5)}$$

Our model addresses all worm interaction factors and can be easily extended to address more types of worms and more number of groups within the network. For example, the basic *SIR* model can also be derived from this model by setting $K_{S^*_1 I_{A1} I_{A2}} = 1$ and $\beta_{11} > 0, S^*_1 > 0, I_{A1} > 0, \gamma > 0$ while setting other parameters to 0.

## IV. EVALUATION

In this paper, we investigate worm interaction and validate our model by using three approaches: (1) model analysis (2) uniform-encounter-based simulation and (3) trace-driven-encounter-based simulation. Our goal is to see the relationships between our proposed model and worm interaction factors. In model analysis, we provide basic conditions that can be used to obtain the metrics. In the uniform-encounter-based simulation, we investigate the effect of worm interaction types, network characteristics and node characteristics on a simple uniform encounter-based network. Then, we evaluate our model on realistic trace-driven-encounter-based simulations. Let us start by analyzing the proposed model.

***A*. Model Analysis**

For brevity, we assume that there are no transitions between groups, i.e., $\lambda_{S*_1S*_2} = \lambda_{S*_2S*_1} = \lambda_{S'_1S'_2} = \lambda_{S'_2S'_1} = \lambda_{I_{A1}I_{A2}} = \lambda_{I_{A2}I_{A1}} = \lambda_{I_{B1}I_{B2}} = \lambda_{I_{B2}I_{B1}} = 0$. We focus our analysis on the aggressive one-sided interaction for two-group encounter-based networks. If we want to suppress the initial infection ($\frac{dI_{A1}}{dt} \le 0$ and $\frac{dI_{A2}}{dt} \le 0$ at $t$=0), from (3-a) and (3-b), then the required conditions for this are

$$S*_1(0)(\beta_{11}I_{A1}(0) + \beta_{12}I_{A2}(0)) \le I_{A1}(0)(\beta_{11}I_{B1}(0) + \beta_{12}I_{B2}(0)) \tag{6-a}$$

$$S*_2(0)(\beta_{22}I_{A2}(0) + \beta_{12}I_{A1}(0)) \le I_{A2}(0)(\beta_{22}I_{B2}(0) + \beta_{12}I_{B1}(0)) \tag{6-b}$$

where $I_{A1}(0), I_{A2}(0), I_{B1}(0), I_{B2}(0)$, $S*_1(0)$ and $S*_2(0)$ are the number of prey-infected nodes, predator-infected nodes and susceptible nodes of group 1 and 2 at $t$=0 respectively.

We obtain from this condition that

$$TI = MI = I_{A1}(0) + I_{A2}(0)\ ,\ I_{A1}(\infty) = I_{A2}(\infty) = 0 \tag{7}$$

where $I_{A1}(\infty)$ and $I_{A2}(\infty)$ are the number of prey-infected nodes of group 1 and 2 at $t=\infty$.

However, we can see from (6-a) and (6-b) that the threshold can only be obtained from such conditions. If those conditions cannot be met, then we can only have certain acceptable level of infection and *TI* can be derived from

$$TI = p\int_{t=0}^{\infty}(S*_2(\beta_{22}I_{A2} + \beta_{12}I_{A1}) + S*_1(\beta_{11}I_{A1} + \beta_{12}I_{A2}))dt \tag{8}$$

*MI* can be found from $(I_{A1} + I_{A2})_{\max}$ where $\frac{dI_{A1}}{dt} = \frac{dI_{A2}}{dt} = 0$ at $t > 0$, in which

$$S*_1(\beta_{11}I_{A1} + \beta_{12}I_{A2}) = I_{A1}(\beta_{11}I_{B1} + \beta_{12}I_{B2}) \tag{9-a}$$

$$S*_2(\beta_{22}I_{A2} + \beta_{12}I_{A1}(0)) = I_{A2}(\beta_{22}I_{B2} + \beta_{12}I_{B1}) \tag{9-b}$$

Because *TL* is the accumulated life of individual prey until the last prey has been removed by predator whose duration indicated by *TR*. we can simply derive *TL* based on the numerical solutions from (3-a) and (3-b) as follows:

$$TL = \sum_{t=o}^{\infty}(I_{A1}(t) + I_{A2}(t))\Delta t \tag{10}$$

Since *AL* is the average lifespan for each node that has been terminated by predator which is equal to the number of nodes that are ever infected, *AL* can be derived from (8) and (10) as

$$AL = \frac{TL}{TI}. \tag{11}$$

We can find $TA$ which is derived from $t$ where $\frac{dS^*_1}{dt} = \frac{dS^*_2}{dt} = \frac{dS'_1}{dt} = \frac{dS'_2}{dt}$ $= \frac{dI_{A1}}{dt} = \frac{dI_{A2}}{dt} = \frac{dI_{B1}}{dt} = \frac{dI_{B2}}{dt} = 0$, $S^*_1(0) = I_{B1}(t)$ and $S^*_2(0) = I_{B2}(t)$ while $TR$ is derived from $t$ where $\frac{dI_{A1}}{dt} = \frac{dI_{A2}}{dt} = 0$, $I_{A1} = I_{A2} = 0$ and $TA \geq TR \geq t_B$ where $t_B$ is the time of last batch arrival.

### *B.* Uniform-encounter-based simulations

We use encounter-level simulations to simulate a simple uniform encounter of 1,000 mobile nodes of a uniform encounter-based network with no batch arrivals and all nodes are susceptible to both prey and predator. Each simulation runs at least 1,000 rounds and we plot the median values for each position. We assume that there is only one group in the network with $\beta$ = 5x10$^{-5}$ sec$^{-1}$ and two groups in part b.3 with $\beta_{11}, \beta_{12}, \beta_{22}$ between $3x10^{-5}$ to $30x10^{-5}$ sec$^{-1}$. In addition, we only assume the aggressive one-sided worm interaction in all parts except in part a.

Before discussing about our simulation results, we need to define the important parameters, initial-infected-node ratio, which we use for uniform-encounter-based simulations. Let $Y$ be an initial-infected-node ratio of predator to prey of the whole network,

$$Y \equiv \frac{\sum_{j=1}^{g} I_{Bj(0)}}{\sum_{j=1}^{g} I_{Aj(0)}} \tag{12}$$

where $g$ is the number of groups in the network and $j$ is the group identification.

Along with our worm interaction factors, $Y$ is used to investigate on the outcomes of having number of initial-predator-infected nodes more than the number of initial-prey-infected nodes within the same networks given that $d$ = 0 (non-zero-delay deployment is investigated in b.3).

#### a. Worm interaction types

As shown in fig. 2, we can clearly see that predator in aggressive one-sided interactions is much more effective than predator in other two worm interaction types for all metrics. Note that we have not shown $TA$ for conservative one-sided and aggressive two-sided worm interaction because $TA = \infty$ and also not shown $TR$, $TL$ and $AL$ for aggressive two-sided worm interaction because $TL = AL = TR = \infty$. Although, *TI, MI, TL, and AL* in the conservative one-sided interaction is at least one order higher than those of aggressive one-sided interaction, but $TR$ in the conservative one-sided interaction is only two-time higher than that of aggressive one-sided interaction (with the same *Y*). This small difference occurs simply because, even with aggressive one-sided interaction, predator infection rate is slow down at the later state of termination/vaccination period. Simplified model for aggressive one-sided, conservative one-sided and aggressive two-sided worm interactions are shown in fig.1b-d, respectively.

Next we focus on the effects of large *Y* on our metrics only with the aggressive one-sided interaction. In fig.3a, *TI* and *MI decrease exponentially as Y increases.* We also find that if $S(0):I_B(0):I_A(0)$ is constant then $MI:N$ and $TI:N$ are also constant even *N* changes. From fig. 3b *TL decreases exponentially as Y increases. AL,* on the other hand, is almost constant for all *Y*. It is interesting to see that *TL* and *AL* are merging at their minimum when $Y = Y_{max}$**. As we can see that $TL_{min}$ and $AL_{min}$ do not reach zero at $Y_{max}$ because the next encounter time of a prey-infected node with *any* of initial predator-infected node ($I_B(0)$) requires $1/I_B(0)\beta$. Furthermore, from (11), $TL_{min} = TI_{min}AL_{min}$, thus $TL_{min}$ and $AL_{min}$ merge to each other because $TI_{min} = I_A(0) = 1$.

From the observation in Fig.3c, *TR* reduces much faster than *TA* with the increase of *Y*. *TR* decreases exponentially as *Y increases*. *TA* starts to be reduced rapidly when $Y \approx Y_{max}$. At $Y_{max}$, we can see that $TA_{min}=TR_{min}=AL_{min}$, Note that *TA* is also similar to the average time for every node to receive a copy of a message from a random source in an encounter-based network which can be derived as $(2\ln N + 0.5772)/N\beta$ [3]. (**$Y_{max}$=1000*)*

**b. Network characteristics**

We start by examining the relationships of the aggressive one-sided interaction and the network characteristics: node size, contact rate and group behavior. For contact rate and node size, we simply assume that the network only have one group to focus only on the effects of these factors to our metrics. After that we would look deeper into the group behaviors including group size, contact rate within a group and the contact rate between groups.

**(1) Network size**: In fig. 4a and b, we find that *TI* and *MI* (as the fraction of *N*) for each *Y* but different *N* are saturated at the same fraction of *N*. Because the fraction of *N* that prey infects susceptible nodes and the fraction of *N* that predator terminates/vaccinates are relatively equivalent for all *N*s. Surprisingly, in fig. 4c, *TL* becomes saturated at certain absolute level and also independent of *N* but depends only on *Y*. This occurs because encounter rate ($\delta$) which is the rate of a node encounters *any node* (i.e., $\delta = \beta(N-1)$) is increasing linearly with *N* (because $\beta$ is fixed, but the number of pairs *N-1* increases as *N* increases) and causes linear reduction of the time between encounter causing *AL* to be reduced proportionally to *N* (as shown in fig. 4d) while *TI is also* increased proportionally to *N* (as shown in fig.4a). The product of these two numbers yields the constant *TL*. In fig. 4e and f, the impact of *N* on *TA* and *TR* is quite similar to *AL*. It is interesting to see that for *Y* = 1 (1:1), *TA* = *TR* for all *N*, and hence it implies that time to remove all preys are simply the time that predator needs to infect and remove prey from all nodes (when *Y* = 1). In sum, we can see that *N* linearly increases *TI* and *MI* and exponentially reduce *AL*, *TA* and *TR*. The effects of $N_n$ (group size) are further investigated in part c.3.

**(2) Contact rate:** As shown in fig. 5a and b, as expected, *TI* and *MI* for each *Y* are relatively constant even with the increase of $\beta$ (because of the equal change of $dI_A/dt$ and $dI_B/dt$). Similar to *N,* as the $\delta$ increases (fixed number of pairs *N*-1, but $\beta$ increases), $\beta$ exponentially decreases *AL*, *TA* and *TR*. However, unlike *N*, *TL* is reduced exponentially as $\beta$ increases, simply because *TI* is constant for all $\beta$. In addition, the lower the *Y,* the higher impact caused by $\beta$ will be. The effects of contact rate of multiple groups are examined next.

**(3) Group behavior:** Earlier, we only assume single-group behavior in a network; in this part we will discuss the two-group behavior. Here we look into effect of the group size, contact rate of one of the two groups, and contact rate between two groups on the worm interactions.

We start by investigating the effects of group sizes as the fraction of fixed *N* (1000 nodes) where $\beta_{11} = 6x10^{-5}\sec^{-1}, \beta_{22} = 9x10^{-5}\sec^{-1}$ and $\beta_{12} = 3x10^{-5}\sec^{-1}$. Group 1 and group 2 are called "slow group" and "fast group", respectively. For the first part (fig. 6a, b and c), an initial prey-infected node is in the slow group and an initial predator-infected node is in the fast group (slow-prey-fast-predator case). In the second part (fig. 6d, e and f), an initial prey-infected host is in the fast group and an initial predator-infected node is in the slow group (fast-prey-slow-predator case).

Here in fig.6a and d, we see that as the size of the *fast* group *increases*, *TI*, *MI*, and *TL* linearly *decrease*. This indicates the independent of which group has the initial predator-infected node or the initial prey-infected node. As *TI* and *TL* linearly decrease with the same rate as of the increase of fast-group size, then *AL* is almost constant for all group sizes. *TA* and *TR* increase gradually as the slow-group size increases (and fast-group size decreases), and drop gradually after reaching their peak value. This occurs because of the low contact rate between groups.

In fig. 7, we show the impact of the contact rate of the initial-prey-infected-node group where the contact rate of initial prey group $\beta_{11} = 3$ to 30x10$^{-5}$ sec$^{-1}$ and the contact rate of the initial predator group $\beta_{22} = 15$x10$^{-5}$ sec$^{-1}$ and contact rate between group $\beta_{12} = 3$x10$^{-5}$ sec$^{-1}$. As expected, *TI*, *MI* and *TL* increase linearly as $\beta_{11}$ increases while *TA* and *TR* reduce exponentially as $\beta_{11}$ increases. This effect is similar to the increase of contact rate in a single group (fig. 5e-f).

In fig. 8, we show the impact of the contact between groups where $\beta_{11} = 3\text{x}10^{-5}$ sec$^{-1}$ and $\beta_{22} = 15\text{x}10^{-5}$ sec$^{-1}$ and $\beta_{12} = 3$ to $30\text{x}10^{-5}$ sec$^{-1}$.As shown in fig. 8a-b, as $\beta_{12}$ increases, prey in the slow-prey-fast-predator can infects more susceptible nodes and predator in the fast-prey-slow-predator can terminate more preys and vaccinate more susceptible nodes (as indicated by *TI* and *MI*). Hence, the *contact rate between groups* only helps *prey or predator in the slower group* to infect relatively more nodes than the one in the faster group (i.e., worms in both groups infects nodes faster but the one in slower group has higher relative improvement). However, *TL*, *AL*, *TA* and *TR* reduce as the contact rate between group increases for all cases (slow-prey-fast-predator and fast-prey-slow-predator cases), and that because $\delta$ increases. We evaluate group characteristics again in trace-driven encounter-based networks (Section C).

**c. Node characteristics**

We vary cooperation (*c*) from 20% to 100%, immunization (*i*) from 0% to 90% with 100% "*on*" time for the first part of experiments (fig. 9a-f) and we vary "*on*" time from 10% to 90% with 90% cooperation and 10% immunization, for the second part (fig.9g-h). The first part aims to analyze the impact of cooperation and immunization whereas the second part aims to analyze the *on-off* behavior on aggressive one-sided worm interaction. In this simulation, again we assume only a single group within the network. Simplified node-characteristic-based aggressive one-sided interaction is shown in fig.1e.

**(1) Cooperation:** In fig. 9a-f, we find that cooperation, surprisingly, reduces prey infection for every metric. (Note that cooperation actually increases absolute *TI* and absolute *MI*, but relative *TI* (or *TI*/ $N*$) and relative *MI* (or *MI*/ $N*$) are reduced where the number of cooperative-susceptible nodes $N* = c(1-i)N$ ). *We can observe that cooperation reduces AL, TA* and *TR significantly* more than it does to *TI*, *MI* and *TL*.

**(2) Immunization:** Similarly, for immunization fig. 9a-f shows that immunization reduces all categories of metrics except *TA* and *AL*. With the increase of immunization, *TI* is reduced much faster than *TL,* thus increase of immunization increases *AL*. Furthermore, increase of immunization, as expected, reduces *TR* because of less number of possible prey-infected nodes.

*Immunization reduces relative TI, relative MI and TL more significantly* than it does other *TR*. With equal increase (20% to 80%), immunization at cooperation = 100% reduces relative *TI*, relative *MI* and *TL* approximately 8.8 times, 2.7 times, and 10.6 times ,respectively, more than cooperation does at immunization = 0%. On the other hand, cooperation reduces *TR* approximately 3.3 times more than immunization does.

As shown in fig. 9e, unlike cooperation, immunization *cannot* reduce *TA*.

**(3) *On-off* behavior:** The impact of *on-off* behavior (*p*) is clear in fig. 9g-h. As expected, with variant of *"on" time*, relative *TI* and relative *MI do not change*. The ratio of contact rate between predator and prey is an indicator of the fraction of infected nodes irrespective of the contact rate. In this case, the ratio of contact rate is always 1.0, and hence the constant of relative *TI* and relative *MI*. Because of the increase of *"on" time* causing reduction of time between consecutive encounters between nodes, hence *TL*, *AL TA* and *TR exponentially decrease as p increases*.

**(4) Delay:** As shown in fig.9i, the delay *(d)* causes absolute *TI* and absolute *MI* to linearly increase until the number of prey-infected node reaches the *N*. Similarly, in fig.9k, *TA* and *TR* also increase linearly as *d* increases. We can notice that the increase of *TA* and *TR* is simply the delay. In addition, *TA* and *TR* are merging after certain delay. For *TL* and *AL* (fig.9j), they slowly increase as *d* increases.

Next, we will apply what we learn from the simulation of worm interaction in the uniform-encounter-based networks to realistic non-uniform encounter-based networks.

### *C.* Trace-driven encounter-based simulations

We investigate the consistency of the model-based results with those generated by using measurement-based real encounters. We drive our encounter-level simulations using the wireless network traces of the University of Southern California of 62 days in spring 2006 semester [5]. We define an encounter as two nodes sharing the same access point at the same time. We randomly choose 1,000 random nodes from 5,000 most active nodes based on their online time from the trace. Their median $\beta$ is 1.27x10$^{-6}$ sec$^{-1}$and median number of unique encounter node is 94. We use $I_A(0)=1$ and $I_B(d)=1$ where *d* is the delay between initial predator-infected node and initial prey-infected node in the simulation. This delay was introduced as the traced delay between the first arrival of two groups which the initial predator-infected node and the initial prey-infected node are assumed to be in different groups (and different batch arrivals). First group and second group account approximately for 90% and 10% of total population, respectively. The first group has average contact rate $\beta_{11}$=3.6x10$^{-6}$ sec$^{-1}$, the second group has average contact rate $\beta_{22}$=3.3x10$^{-6}$ sec$^{-1}$, and the approximated contact rate between group $\beta_{12}$=4x10$^{-7}$ sec$^{-1}$. When contact rate of the initial predator-infected node is higher than that of the initial prey-infected node, we call this scenario "*Fast predator*". On the other hand, when contact rate of initial predator-infected node is lower than that of prey, we call this scenario "*Slow* *predator*". From the trace, the median arrival delay between initial predator-infected node and initial prey-infected node is 8.7 days (introduced by the gap between the first and the second batch arrivals). Because the first group is in the first batch, hence "*Fast predator*" is also the *early* predator and "Slow predator" is also the *late* predator.

We can see the consistent batch arrival pattern in fig. 7c, each line represents different start new-node arrival time into the networks, i.e., day 0, 10, 20 and 30 where day 0 is January 25, 2006. Because at the beginning of the semester, not all students had returned to campus; hence, the large gap between batch arrivals existed. The smaller gaps (1 day) in other start days were caused by the university's schedule that has classes either on Tuesday-Thursday or Monday-Wednesday-Friday. *Hence, the batch arrival patterns are likely to occur in any encounter-based networks due to the users' schedules*. In addition, in fig. 10a-b, we find that user's encounter in the trace is highly skewed (non-uniform), i.e., top 20% of user's total encounter account for 72% of all users' encounters and 70% of users encounter less than 20% of total unique users which are caused by non-uniform *on-off behavior and location preferences* [5, 6].

We choose to run our trace-driven simulations at day 0 to see the significance of batch arrival patterns on worm interactions. To validate our model accuracy, we compare the trace-driven simulation results with our aggressive one-sided model with node characteristics and group behavior. We also apply the batch arrival and delay to our model and compare the trace-driven simulation results with our model plot.

In our model, we use $\beta_{11} = 3.6x10^{-6}, \beta_{22} = 3.3x10^{-6}, \beta_{12} = 4x10^{-7}$ with $t_1$= day 8.7 (second batch arrival, 395 nodes join group 1, 50 nodes join group 2), $t_2$ = day 8.71 (all predator-infected nodes leaving the networks), $t_3$ = day 11.57 (predator-infected nodes rejoin the networks), $t_4$ = day 17.4 (third batch arrival, 50 nodes join group 2), $t_4$ = day 40.5 (fourth batch arrival, 5 nodes join group 2). These batch arrival patterns are approximated from the observed trace and simulations.

In fig. 10f-i and l-o, this batch arrival patterns and the delay cause significant additions on our proposed metrics especially *TL*, *AL*, *TA*, and *TR* (*TA* is subject to the time of the last-node arrival). In addition, we find that immunization (*i*) is still a very important factor to reduce relative *TI*, relative *MI*, *TL*, and *TR,* in the *"Slow predator"* case, but it does not have much impact in the "*Fast predator*" case, since there is not much room for improvement (except *TL*). However, unlike uniform-encounter worm interaction, we find that *cooperation only helps reduce relative TI, relative MI, TL, AL and TR in "Fast predator" case*.

In fig. 10d-f, relative *TI,* relative *MI, TL* with "*Slow predator*" almost linearly decrease to zero with the increase of *i*. Hence*, large immunization can offset large delay*. Surprisingly, as shown in fig. 10g and m, *AL* with "*Fast predator*" has not shown significant improvement over *AL* with "*Slow predator*".

Our model seems to more accurately predict the metrics in "*Slow predator*" case in which delay and batch arrival pattern are the major factors. On the other hands, for the "*Fast predator", TI* and *MI* (fig.10j-k) are more sensitive to fine-grained non-uniform encounter patterns in which we simplify them to only two-group encounters. With the number of groups precisely estimated, the accuracy of the metrics estimations can be drastically improved.

## V. SUMMARY AND FUTURE WORK

In this paper, we propose a general worm interaction model addressing worm interaction types, network characteristics and node characteristics for encounter-based networks. In addition, new metrics as a performance evaluation framework for worm interactions are proposed. We find that predator is most effective in aggressive one-sided worm interaction. In addition, we find that in uniform and realistic encounter-based networks, immunization and delay are the most influential node characteristics for total prey-infected nodes, maximum prey-infected nodes and total prey lifespan. Cooperation and *on-off* behaviors greatly affect average individual prey lifespan, time to secure all nodes and time to remove all preys in uniform encounter-based networks. Furthermore, for multi-group uniform-encounter-based networks, large group-size with fast contact rate helps limit total prey-infected nodes, maximum prey-infected nodes. Fast contact rate between groups reduces average individual prey lifespan, time to secure all nodes and time to remove all preys. Our model shows a very good agreement with uniform-encounter simulation results.

Based on realistic mobile networks measurements, we find that batch arrivals are common in the trace and likely to take place in any encounter-based networks. In addition, we also find that the contact rate and the number of unique encounters of users are highly skewed. This network characteristic causes worm infection behavior to deviate from our predictions, even though the general trends remain similar to the model. We believe that our general worm interaction model can be extended to incorporate fined-grained and dynamic user groups to enhance the accuracy of prediction.

In such networks, immunization and timely predator deployment seems to be more important factors than cooperation. Hence, enforcing early immunization and having a mechanism to identify a high-contact-rate group to deploy an initial predator-infected node is critical to contain worm propagation in encounter-based networks. These findings provide insight that we hope would aid to develop counter-worm protocols in future encounter-based networks.

## References

[1] F. Castaneda, E.C. Sezer, J. Xu, *"WORM vs. WORM: preliminary study of an active counter-attack mechanism"*, ACM workshop on Rapid malcode, 2004

[2] Z. Chen, L. Gao, and K. Kwiat, "*Modeling the Spread of Active Worms*", IEEE INFOCOM 2003

[3] D.E. Cooper, P. Ezhilchelvan, and I. Mitrani, *A Family of Encounter-Based Broadcast Protocols for Mobile Ad-hoc Networks*, In Proceedings of the Wireless Systems and Mobility in Next Generation Internet. 1st International Workshop of the EURO-NGI Network of Excellence, Dagstuhl Castle, Germany, June 7-9 2004

[4] W. Hsu, A. Helmy, "*On Nodal Encounter Patterns in Wireless LAN Traces*", The 2nd IEEE Int.l Workshop on Wireless Network Measurement (WiNMee), April 2006.

[5] W. Hsu, A. Helmy, "*On Modeling User Associations in Wireless LAN Traces on University Campuses*", The 2nd IEEE Int.l Workshop on Wireless Network Measurement (WiNMee), April 2006.

[6] A. Ganesh, L. Massoulie and D. Towsley, *The Effect of Network Topology on the Spread of Epidemics*, in IEEE INFOCOM 2005.

[7] W. O. Kermack and A. G. McKendrick: "*A Contribution to the Mathematical Theory of Epidemics"*. Proceedings of the Royal Society 1997; A115: 700-721.

[8] D. Moore, C. Shannon, G. M. Voelker, and S. Savage, "*Internet Quarantine: Requirements for Containing Self Propagating Code*", in IEEE INFOCOM 2003.

[9] D. M. Nicol, "*Models and Analysis of Active Worm Defense*", Proceeding of Mathematical Methods, Models and Architecture for Computer Networks Security Workshop 2005.

[10] P. Szor, "*The Art of Computer Virus Research and Defense*" (Symantec Press) 2005

[11] S. Tanachaiwiwat, A. Helmy, "*Worm Ecology in Encounter-based Networks (Invited Paper)*" *IEEE Broadnets* 2007

[12] S.Tanachaiwiwat, A. Helmy, "*On the Performance Evaluation of Encounter-based Worm Interactions Based on Node Characteristics*" Technical Report arXiv:0706.2025

[13] Trend Micro Annual Virus Report 2004 http://www.trendmicro.com

[14] H. Trottier and P. Phillippe, "*Deterministic Modeling Of Infectious Diseases: Theory And Methods*" The Internet Journal of Infectious Diseases ISSN: 1528-8366

[15] A.Vahdat and D. Becker. *Epidemic routing for partially connected ad hoc networks*. Technical Report CS-2000.

[16] M. Vojnovic and A. J. Ganesh, *"On the Effectiveness of Automatic Patching"* , ACM WORM 2005, The 3rd Workshop on Rapid Malcode, George Mason University, Fairfax, VA, USA, Nov 11, 2005.

[17] X. Zhang, G. Neglia, J. Kurose, and D. Towsley. "*Performance Modeling of Epidemic Routing*", to appear Elsevier Computer Networks journal, 2007

[18] C. C. Zou, W. Gong and D. Towsley, "*Code red worm propagation modeling and analysis*" Proceedings of the 9th ACM CCS 2002

Table I. Parameters and definitions

| Parameter | Definition |
|---|---|
| $S, S_n$ | Susceptible nodes: the number of nodes of the whole population that can be infected by either prey or predator, the number of susceptible nodes of group *n* |
| $S^*_n, S'_n$ | Number of susceptible nodes of group *n* that can be infected by either prey or predator, the number of susceptible nodes of group *n* that can be infected by predator only |
| $I_A$ , $I_B$ | Prey-infected nodes: the number of nodes infected by prey of a whole population, Predator-infected nodes: the number of nodes infected by predator of a whole population |
| $I_{An}$ , $I_{Bn}$ | Prey-infected nodes: the number of nodes infected by prey in group *n*, Predator-infected nodes: the number of nodes infected by predator in group *n* |
| $N, N^*, N_n$ | Total number of vulnerable nodes in the networks: it is the sum of number of susceptible nodes, prey-infected nodes and predator-infected nodes, total number of cooperative-susceptible nodes of a whole population, total number of vulnerable nodes of group *n* |
| $\beta$ , $\beta_{nm}$ | Pair-wise contact rate: a frequency of a pair of nodes make a contact with each other of a whole population, a contact rate between a member in group *n* and a member in group *m*. |
| $\delta$ | Encounter rate: a frequency of a node encounters any other node in the same network |
| $Y$ | Initial-infected-nodes ratio: a ratio between predator-infected nodes and prey-infected nodes of the whole population at $t = 0$. |
| $c$ | Cooperation: node's willingness to forward messages for others of the whole population (fraction) |
| $i$ | Immunization: immune nodes (fraction) of the whole population will not be infected by prey |
| $p$ | On-off behavior: "*on*" nodes can participate in forwarding packets while "*off*" nodes cannot (probability) |
| $d$ | Delay: the time differences between initial prey-infected nodes and initial predator-infected nodes |
| $a$ | Re-susceptible: infected nodes can become susceptible again |
| $K_{S^*_1 I_{A1} I_{A2}}$ , $K_{S^*_2 I_{A1} I_{A2}}$ , $K_{S^*_1 I_{B1} I_{B2}}$ , $K_{S^*_2 I_{B1} I_{B2}}$ , $K_{S'_1 I_{B1} I_{B2}}$ , $K_{S'_2 I_{B1} I_{B2}}$ , $K_{I_{A1} I_{B1} I_{B2}}$ , $K_{I_{A2} I_{B1} I_{B2}}$ | State transition indicators: the numbers (0 or 1) used to identify the types of worm interaction types |
| $\Delta_{S^*_1}$ , $\Delta_{S^*_2}$ , $\Delta_{S'_1}$ , $\Delta_{S'2}$ | Batch arrival (and departure) rate: a rate of new vulnerable nodes join (or leave) into the networks |
| $\lambda_{S^*_n S^*_m}$ , $\lambda_{S'_n S'_m}$ , $\lambda_{I_{An} I_{Am}}$ , $\lambda_{I_{Bn} I_{Bm}}$ | Group transition rate: rates of susceptible nodes, susceptible nodes which immune to prey, prey-infected nodes, predator-infected nodes in group *n* become susceptible nodes, susceptible nodes which immune to prey, prey-infected nodes, predator-infected nodes in group *m*, respectively |

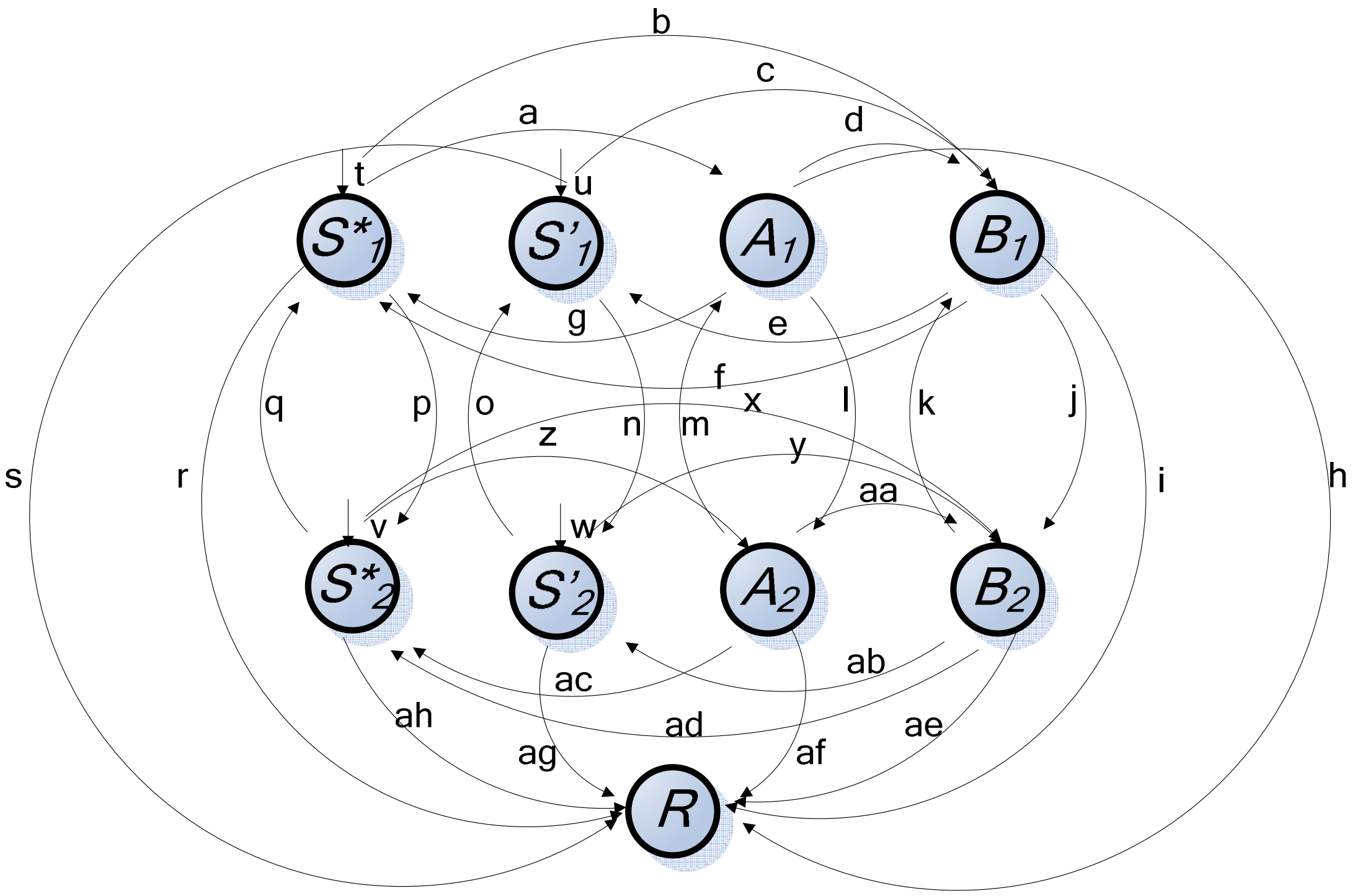

a : $pK_{S^*_1 I_{A1} I_{A2}} S^*_1 (\beta_{11} I_{A1} + \beta_{12} I_{A2})$

b : $pK_{S^*_1 I_{B1} I_{B2}} S^*_1 (\beta_{11} I_{B1} + \beta_{12} I_{B2})$

c : $pK_{S'_1 I_{B1} I_{B2}} S'_1 (\beta_{11} I_{B1} + \beta_{12} I_{B2})$

d : $pK_{I_{A1} I_{B1} I_{B2}} I_{A1} (\beta_{11} I_{B1} + \beta_{12} I_{B2})$

e : $\alpha i I_{B1}$

f : $\alpha(1-i) I_{B1}$

g : $\alpha I_{A1}$

h : $\gamma I_{A1}$

i : $\gamma I_{B1}$

j : $(\lambda_{I_{B1}I_{B2}} - \lambda_{I_{B2}I_{B1}}) I_{B1}$

k : $(\lambda_{I_{B2}I_{B1}} - \lambda_{I_{B1}I_{B2}}) I_{B2}$

l : $(\lambda_{I_{A1}I_{A2}} - \lambda_{I_{A2}I_{A1}}) I_{A1}$

m : $(\lambda_{I_{A2}I_{A1}} - \lambda_{I_{A1}I_{A2}}) I_{A2}$

n : $(\lambda_{S'_1 S'_2} - \lambda_{S'_2 S'_1}) S'_1$

o : $(\lambda_{S'_2 S'_1} - \lambda_{S'_1 S'_2}) S'_2$

p : $(\lambda_{S^*_1 S^*_2} - \lambda_{S^*_2 S^*_1}) S^*_1$

q : $(\lambda_{S^*_2 S^*_1} - \lambda_{S^*_1 S^*_2}) S^*_2$

r : $\gamma_S S^*_1$

s : $\gamma_S S'_1$

t : $\Delta_{S^*_1}$

u : $\Delta_{S'_1}$

v : $\Delta_{S^*_2}$

w : $\Delta_{S'_2}$

x : $pK_{S^*_2 I_{B1} I_{B2}} S^*_2 (\beta_{22} I_{B2} + \beta_{12} I_{B2})$

y : $pK_{S'_2 I_{B1} I_{B2}} S'_2 (\beta_{22} I_{B2} + \beta_{12} I_{B1})$

z : $pK_{S^*_2 I_{A1} I_{A2}} S^*_2 (\beta_{22} I_{A2} + \beta_{12} I_{A1})$

aa : $pK_{I_{A2} I_{B1} I_{B2}} I_{A2} (\beta_{22} I_{B2} + \beta_{12} I_{B1})$

ab : $\alpha i I_{B2}$

ac : $\alpha I_{A2}$

ad : $\alpha(1-i) I_{B2}$

ae : $\gamma I_{B2}$

af : $\gamma I_{A2}$

ag : $\gamma_S S'_2$

ah : $\gamma_S S^*_2$

**Figure 1(a): General worm interaction model state diagram**

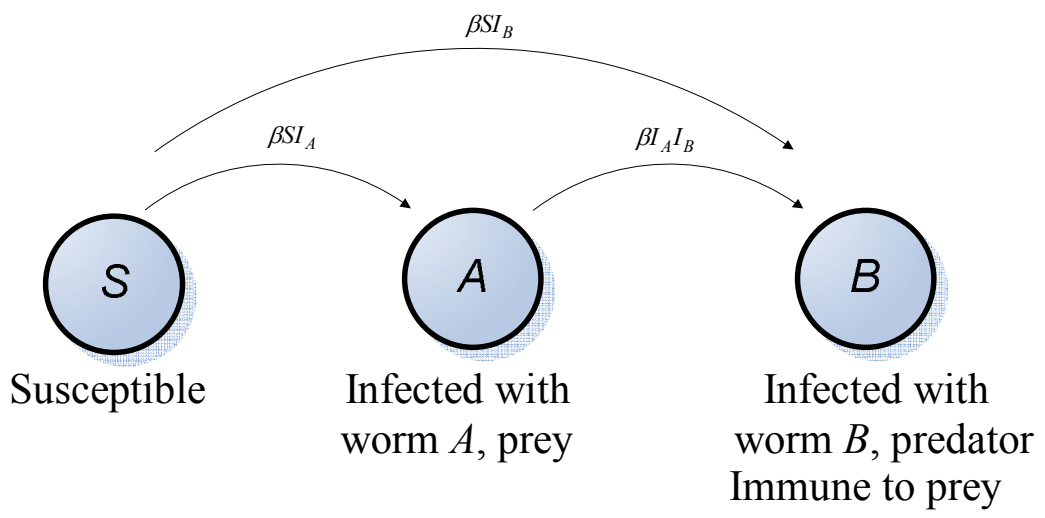

**Figure 1(b): Aggressive one-sided interactions**

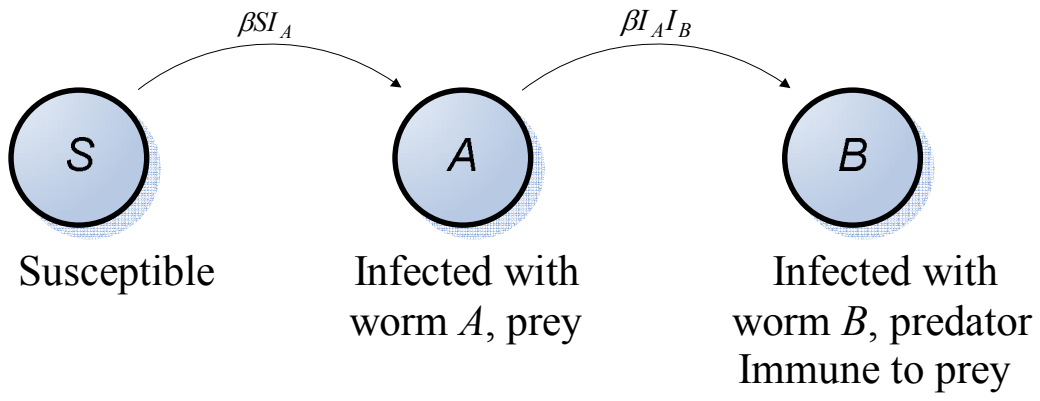

**Figure 1(c): Conservative one-sided interactions**

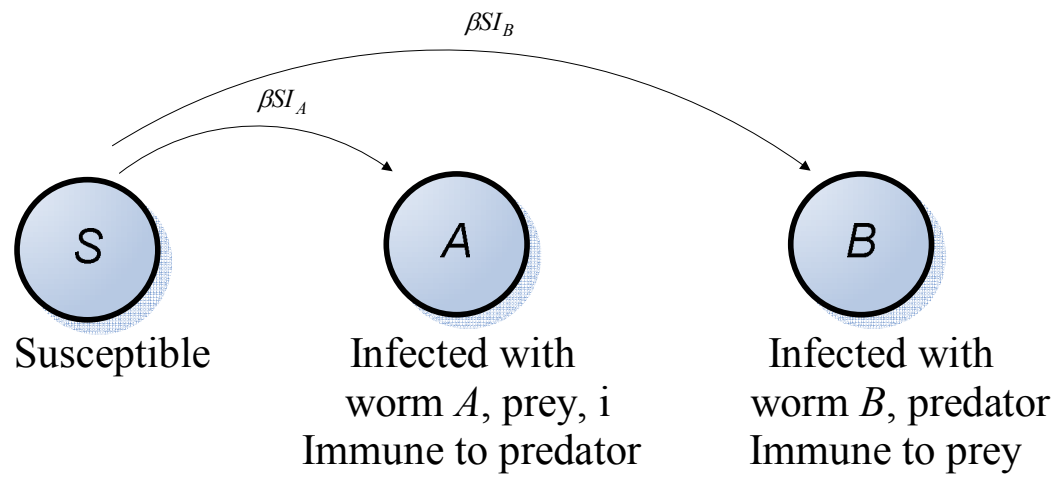

**Figure 1(d): Aggressive Two-sided Interaction**

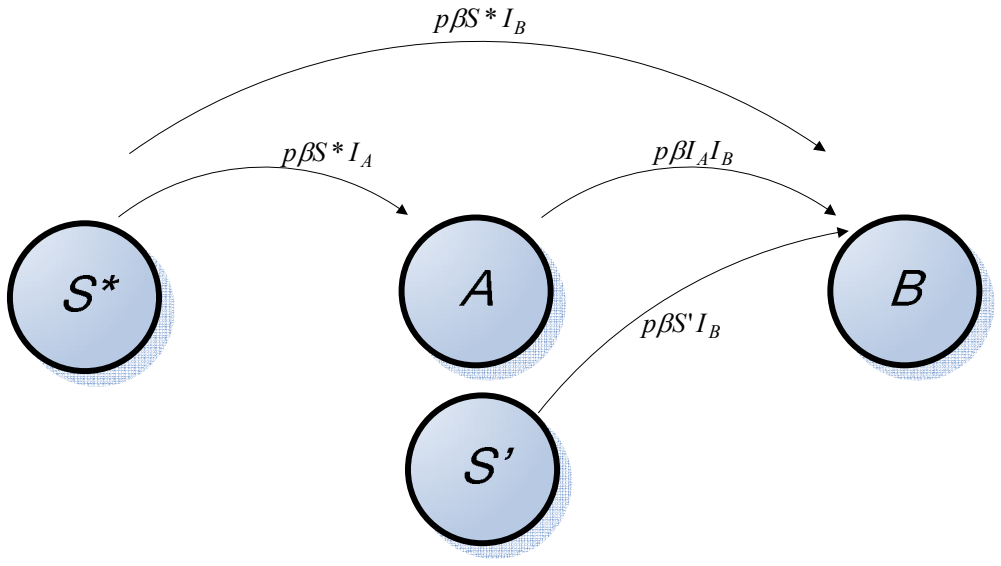

**Figure 1(e): Aggressive one-sided interaction with node characteristics**

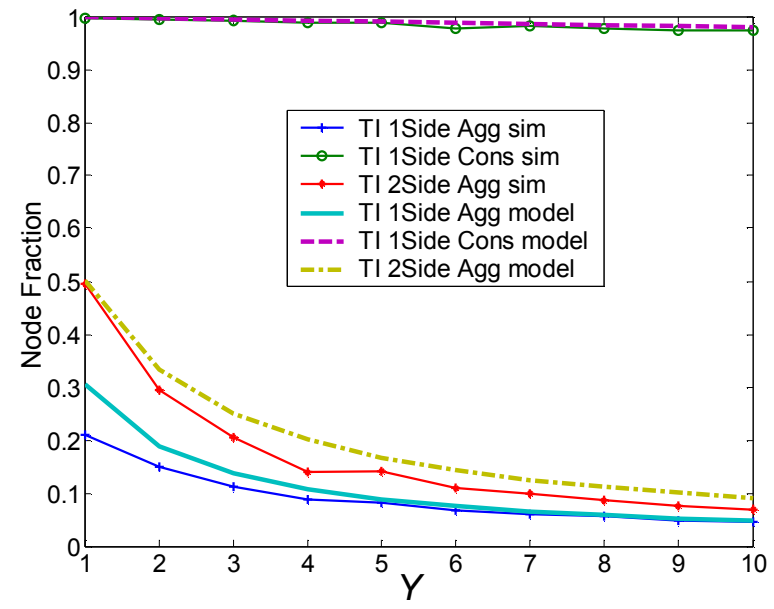

(a)

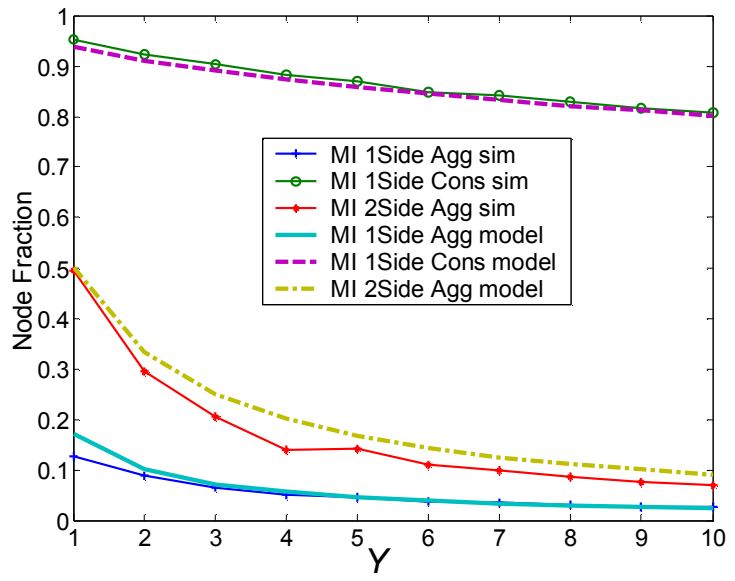

(b)

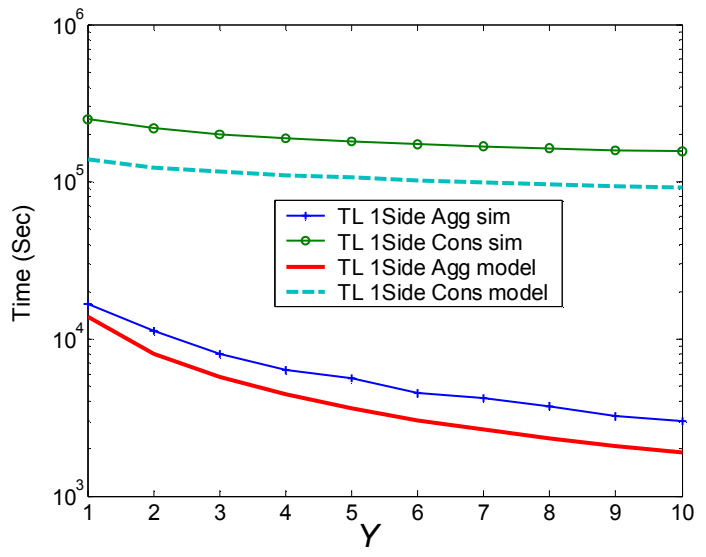

(c)

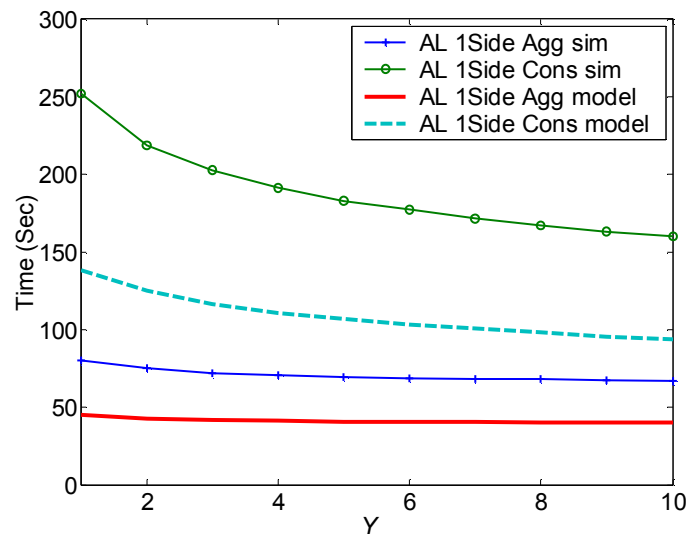

(d)

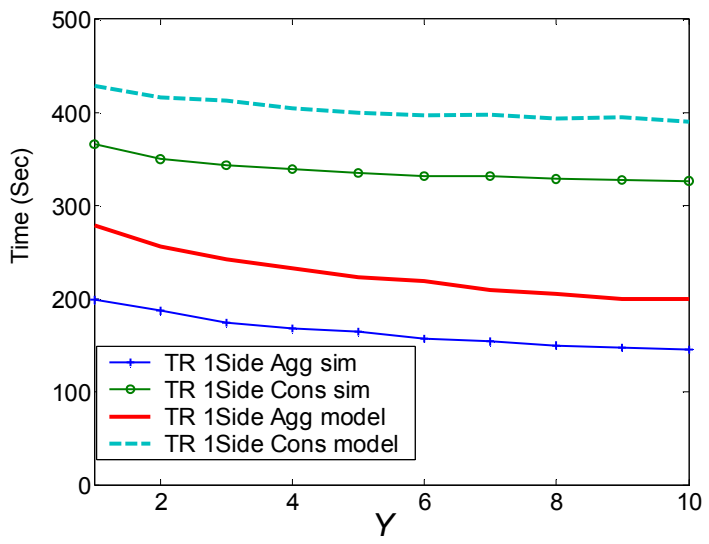

(e)

**Figure 2: Relationships of worm characteristics with *Y***

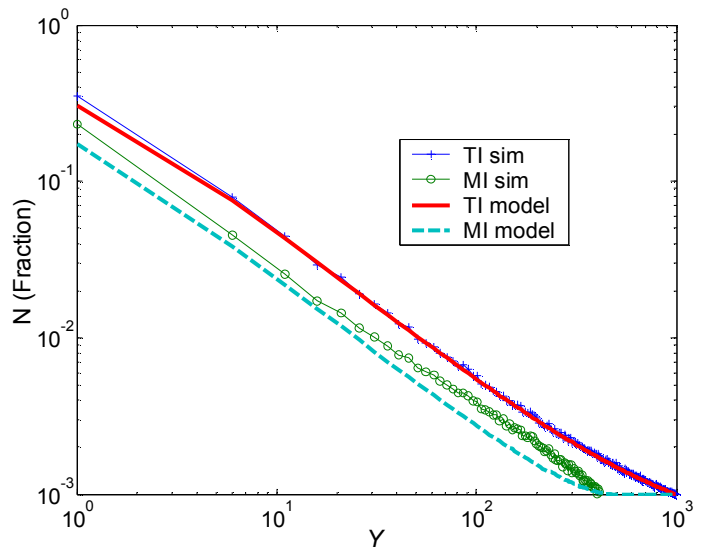

(a)

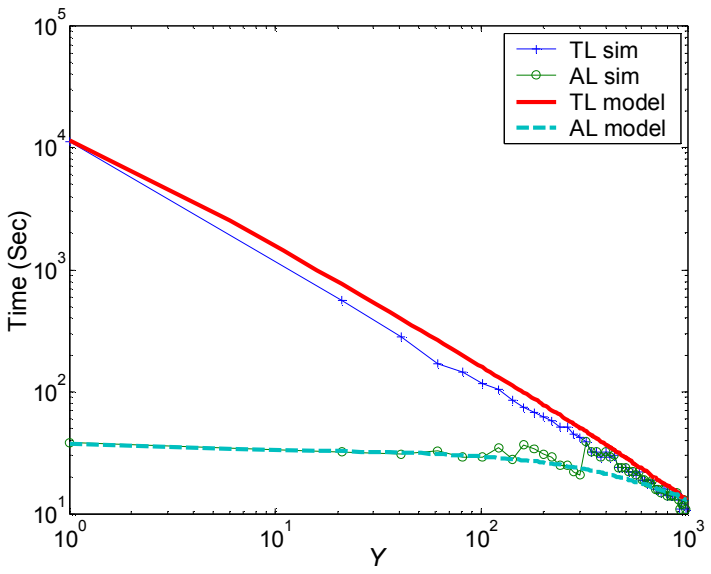

(b)

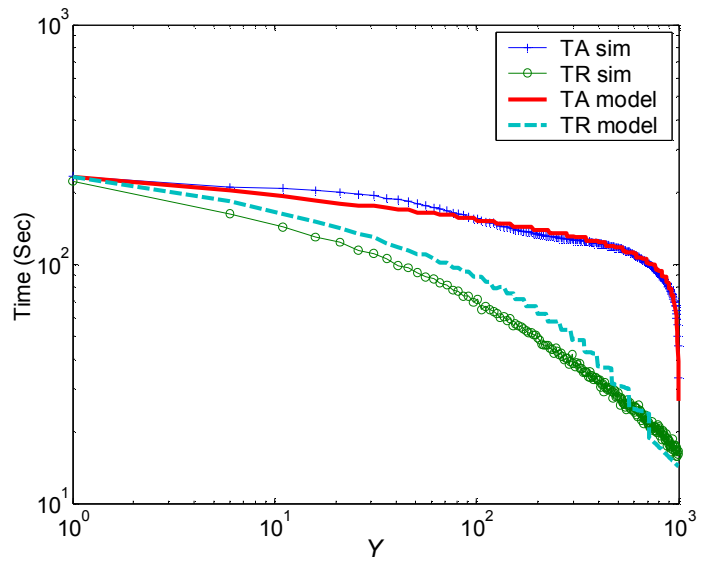

(c)

**Figure 3: Relationships of aggressive one-side interaction with *Y***

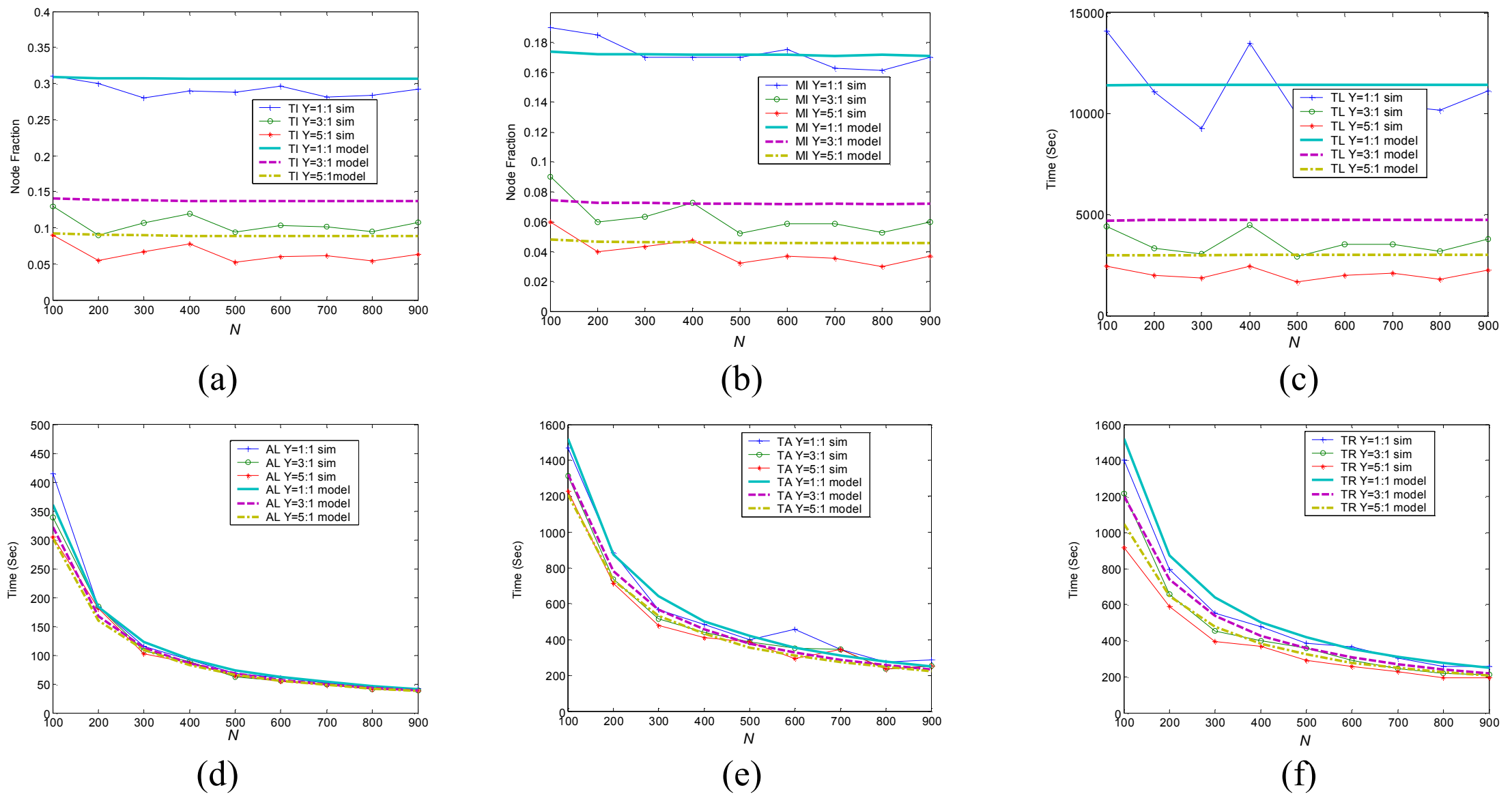

**Fig. 4 Relationships of *N* with metrics**

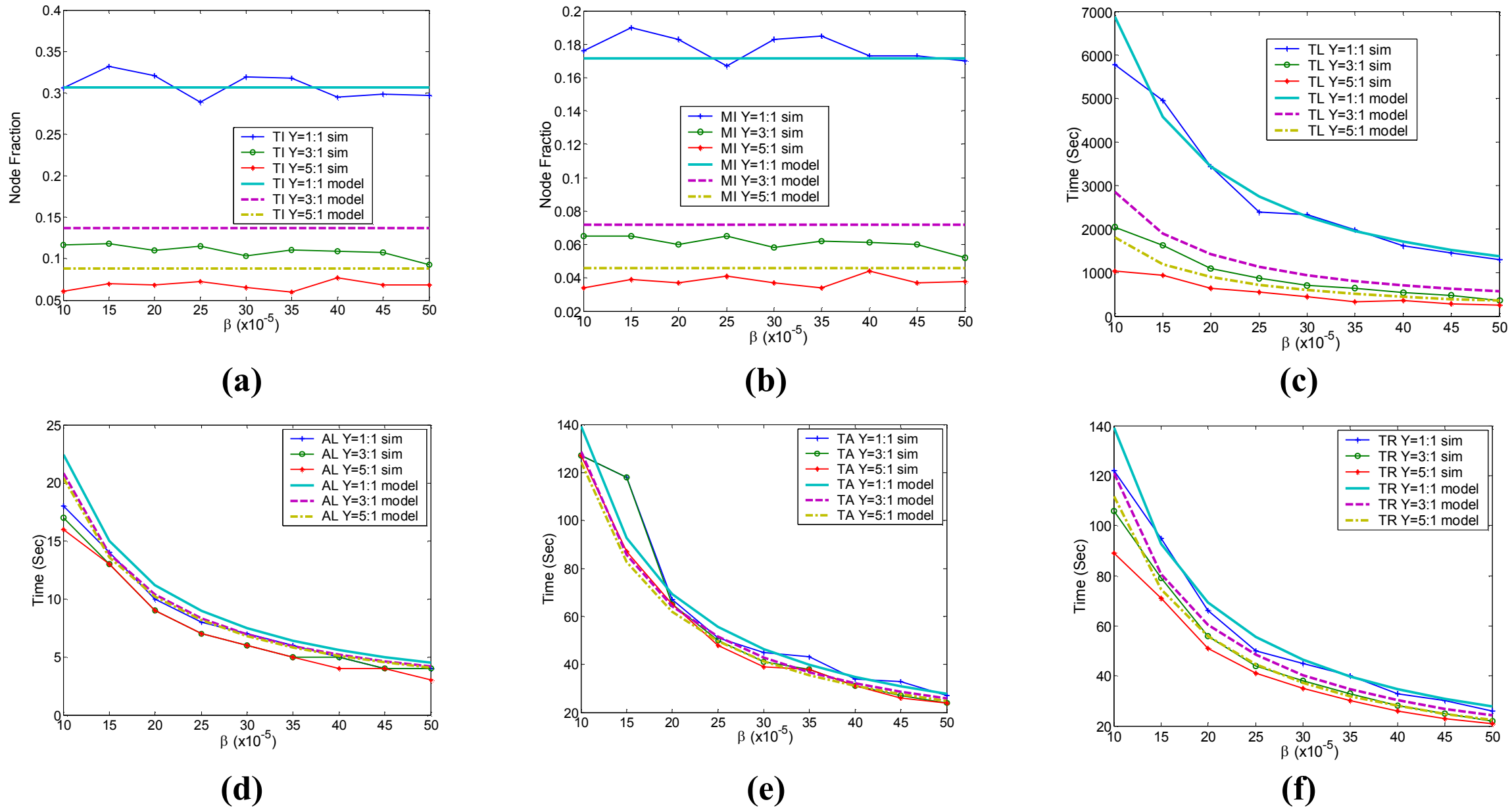

**Fig. 5 Relationships of *β* with metrics**

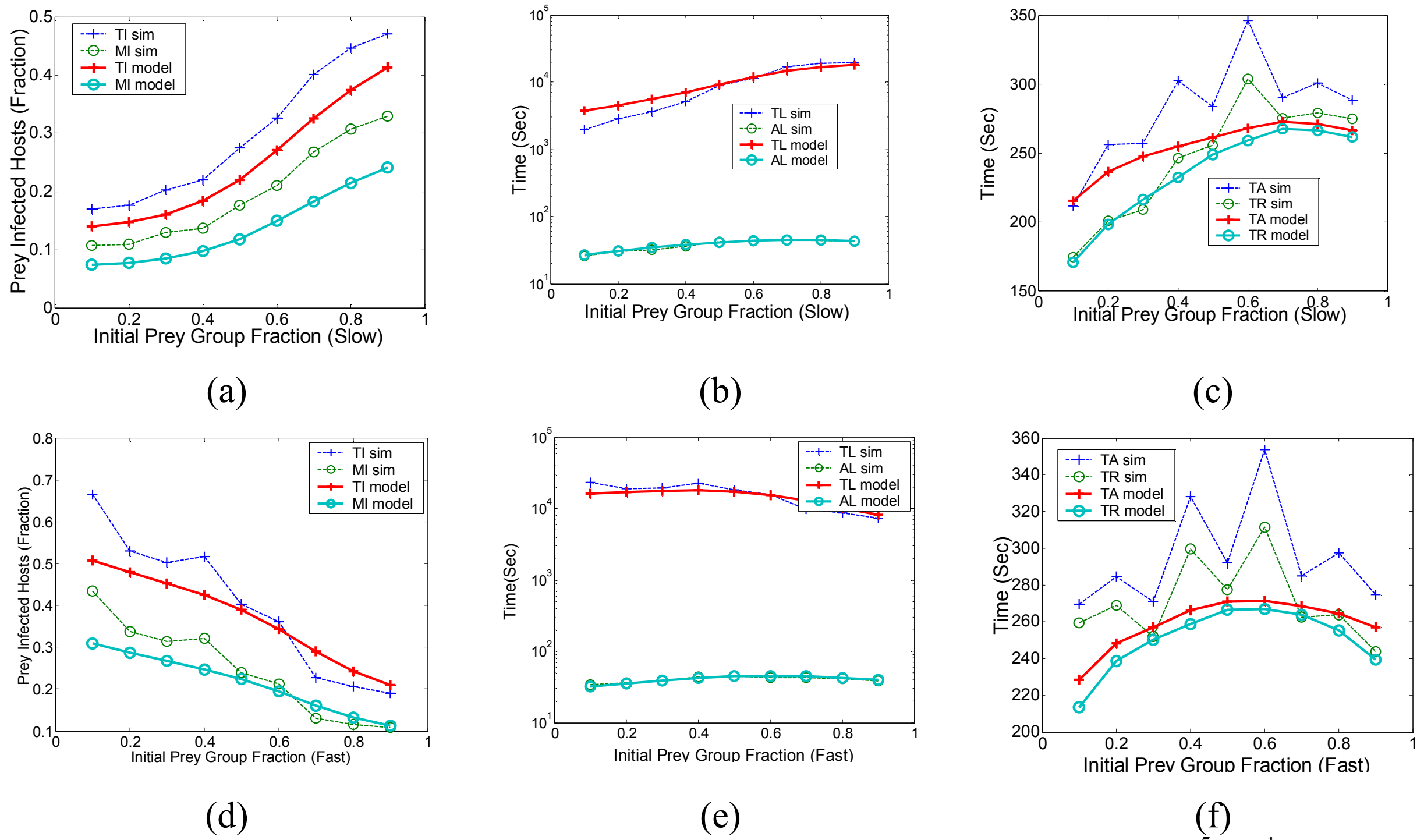

**Figure 6: Effects of group size in two-group population: slow group (contact rate=6x10$^{-5}$ sec$^{-1}$) and fast groups (contact rate= 9x10$^{-5}$ sec$^{-1}$ and contact rate between group =3x10$^{-5}$ sec$^{-1}$) for Slow prey Fast predator (a, c and e) and Fast prey Slow predator (b, d and f)**

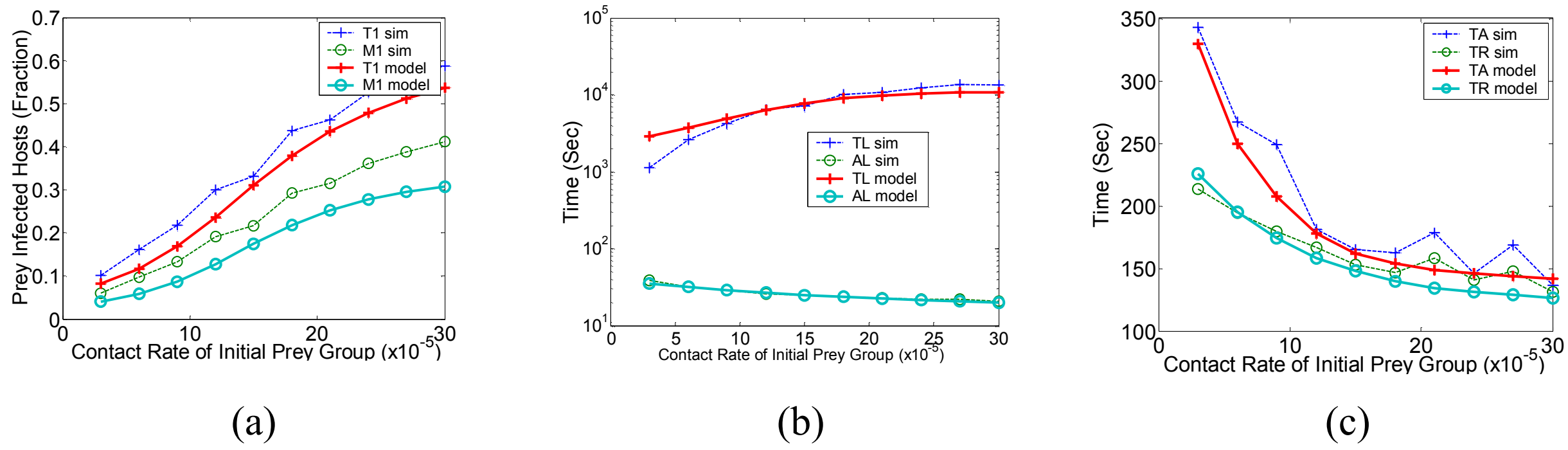

**Figure 7: Effects of initial-prey-infected-node group's contact rate in two group population: varied-contact-rate of initial-prey-infected-node group (contact rate=3 to 30x10$^{-5}$ sec$^{-1}$) and fixed-contact-rate of initial predator group (contact rate= 15x10$^{-5}$ sec$^{-1}$ and contact rate between group =3x10$^{-5}$ sec$^{-1}$)**

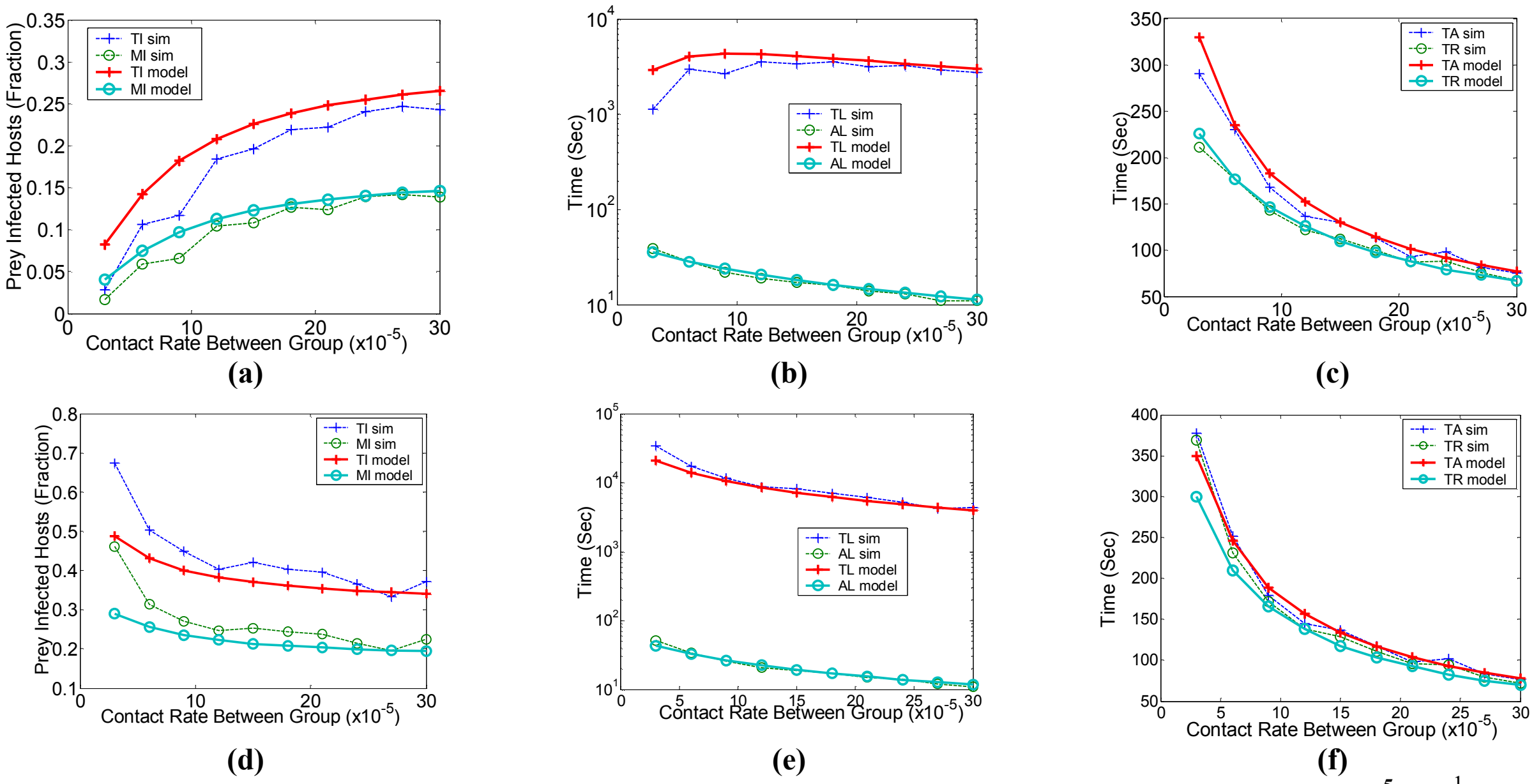

**Figure 8: Effects of contact rate between groups of two-group population: slow group (contact rate=3x10$^{-5}$ sec$^{-1}$) and fast encountered groups (contact rate= 15x10$^{-5}$ sec$^{-1}$and contact rate between group =3 to 30x10$^{-5}$ sec$^{-1}$) for Slow prey Fast predator (a, c and e) and Fast prey Slow predator (b, d and f)**

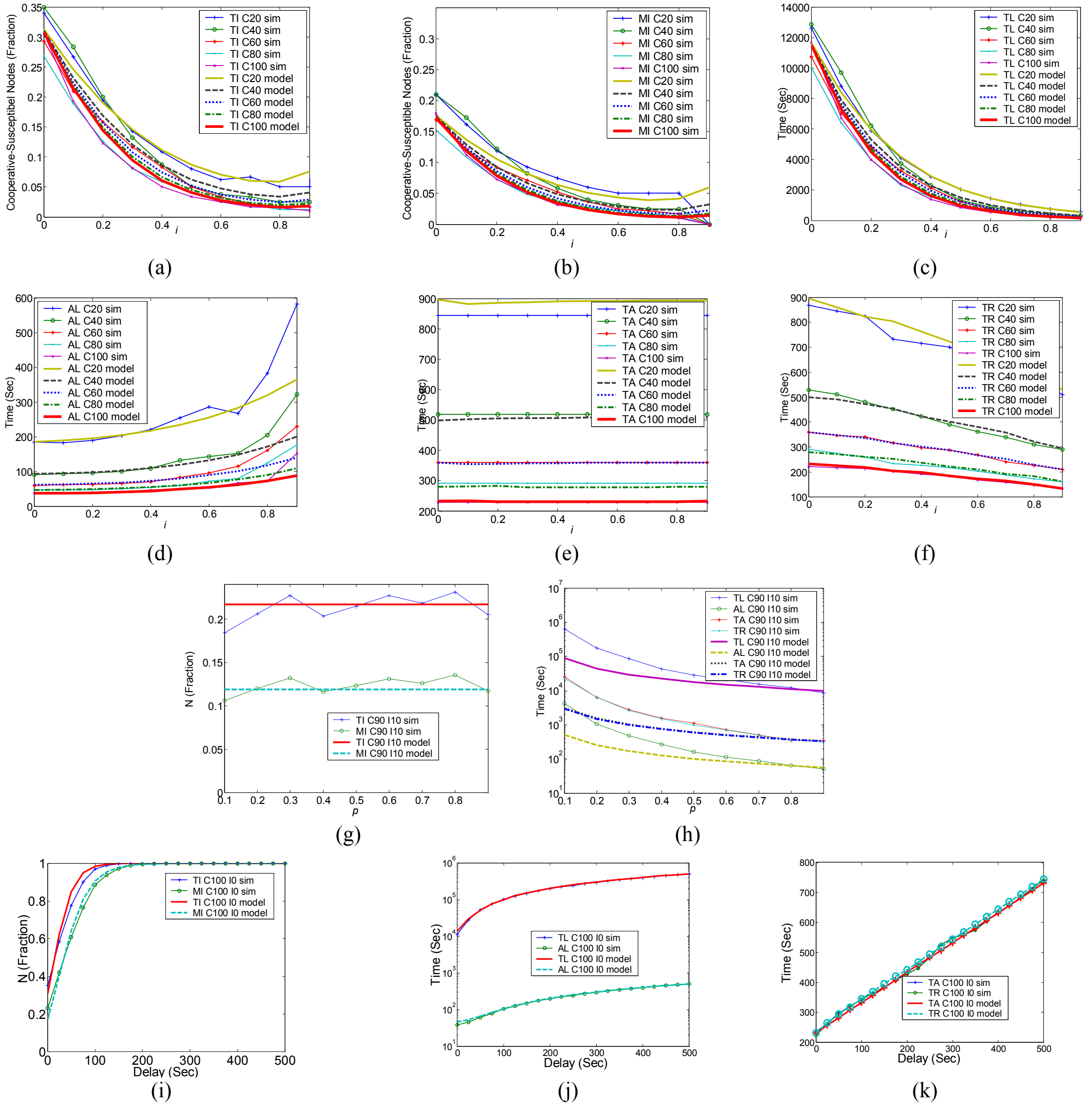

**Figure 9: Effects of cooperation (*c*), immunization (*i*), *on-off* behavior (*p*) and delay (*d*) on uniform-encounter worm interactions**

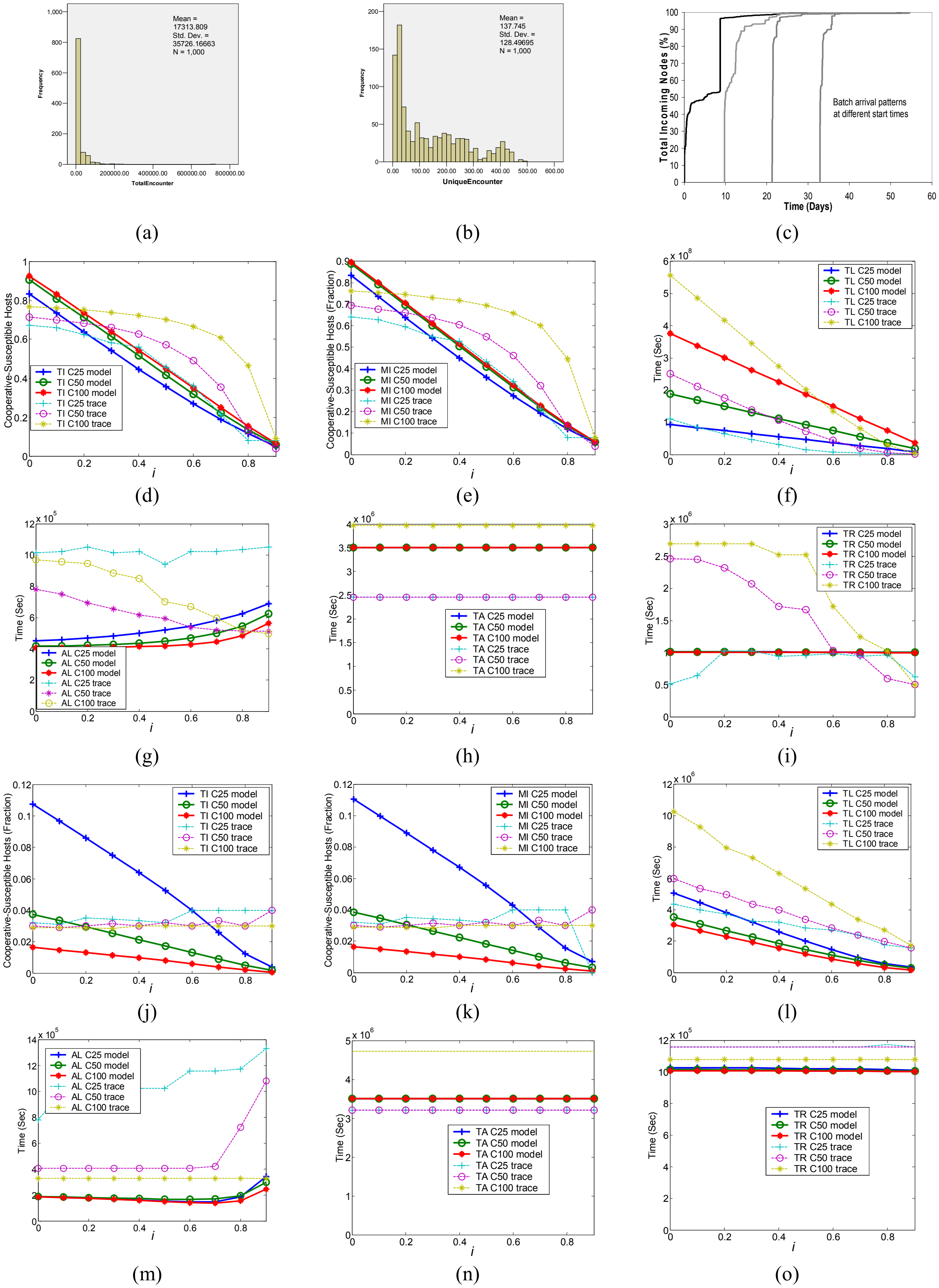

**Figure 10: Trace-based statistics and simulation results: histograms of (a) total encounter/node, (b) unique encounter/node and (c) batch arrival pattern, and effects on cooperation (*c*), and immunization (*i*) on *TI* , *MI* ,*TL*, *AL* , *TA* and *TR* in non-uniform-encounter worm interaction which (d)-(i) initial predator-infected hosts in slow contact-rate and late group, (j)-(o) initial predator-infected hosts in fast contact-rate and early group**